\documentclass[aps,amsmath,amssymb,amsthm,dsfont,prx,nofootinbib,twocolumn,superscriptaddress,nobibnotes]{revtex4-2}
\usepackage{times}
\usepackage[colorlinks=true,citecolor=blue,linkcolor=blue]{hyperref}
\usepackage{graphicx,epstopdf}              
\usepackage{mathtools}                        
\usepackage{amsfonts,amsthm,amssymb,latexsym,amsmath}
\usepackage{mathrsfs}
\usepackage{bbm}
\usepackage{units}  
\usepackage{textcomp} 
\usepackage{hhline}
\usepackage{multirow}   
\usepackage{makecell}
\usepackage{tabularx}
\usepackage{booktabs}
\usepackage{hyperref}
\usepackage{diagbox}
\hypersetup{
colorlinks = true,
linkcolor=magenta,
citecolor=[rgb]{0.15,0.65,0.13},
urlcolor = [rgb]{0.25, 0.41, 0.88}}
\usepackage{float}
\usepackage{cleveref}
\usepackage{tikz}
\usepackage{booktabs}
\usepackage{cancel}
\usepackage{colortbl}
\usetikzlibrary{decorations.markings}

\def\bra#1{\mathinner{\langle{#1}|}}
\def\ket#1{\mathinner{|{#1}\rangle}}
\def\Ket#1{\mathinner{\left|{#1}\right\rangle}}
\def\Bra#1{\mathinner{\left\langle{#1}\right|}}
\def\inner#1{\mathinner{\langle{#1}\rangle}}
\def\bs#1{\boldsymbol{#1}}

\def\aut#1#2{\sigma^{#1} \left [ {#2} \right]}

\def\ZZ{\mathbb Z}

\def\CC{\mathbb C}
\def\cX{\mathcal X}
\def\cZ{\mathcal Z}
\def\cC{\mathcal C}
\def\lbar{\overline}
\def\l{0.9}
\newcommand{\nablapic}{\raisebox{.0\height}{\includegraphics[scale=0.1]{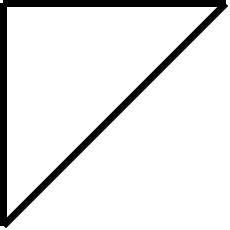}}}

\def\emptycirc{\raisebox{.5pt}{\textcircled{\raisebox{-.9pt} {}}} }

\newcommand{\fs}[2]{{^{#1}}{#2}}

\tikzset{
    vertex/.style={fill,circle,draw,scale=0.3},
    ->/.style={decoration={markings,mark=at position 0.5 with {\fill (2pt,0)--(-2pt,2.31pt)--(-2pt,-2.31pt)--cycle;}},postaction={decorate}},
    ->2/.style={decoration={markings,mark=at position 1 with {\fill (2pt,0)--(-2pt,2.31pt)--(-2pt,-2.31pt)--cycle;}},postaction={decorate}},
}

\makeatletter
\def\l@subsection#1#2{}
\def\l@subsubsection#1#2{}
\makeatother

\begin{document}

\title{Non-Abelian Hybrid Fracton Orders}

\author{Nathanan Tantivasadakarn}
\email{ntantivasadakarn@g.harvard.edu}
\affiliation{Department of Physics, Harvard University, Cambridge, Massachusetts 02138, USA}
\author{Wenjie Ji}
\email{wenjieji@ucsb.edu}
\affiliation{Department of Physics, University of California, Santa Barbara, California 93106, USA}
\author{Sagar Vijay}\email{sagar@physics.ucsb.edu}
\affiliation{Department of Physics, University of California, Santa Barbara, California 93106, USA}

\begin{abstract}
We introduce lattice gauge theories which describe  three-dimensional, gapped quantum phases exhibiting the phenomenology of both conventional three-dimensional topological orders and fracton orders, starting from a finite group $G$, a choice of an Abelian normal subgroup $N$, and a choice of foliation structure. These hybrid fracton orders -- examples of which were introduced in a previous paper \cite{TJV1} -- can also host immobile, point-like excitations that are non-Abelian, and therefore give rise to a protected degeneracy.  We construct solvable lattice models for these orders which interpolate between a conventional, three-dimensional $G$ gauge theory and a pure fracton order, by varying the choice of normal subgroup $N$.  We demonstrate that certain universal data of the topological excitations and their mobilities are directly related to the choice of $G$ and $N$, and also present complementary perspectives on these orders: certain orders may be obtained by gauging a global symmetry which enriches a particular fracton order, by either fractionalizing on or permuting the excitations with restricted mobility, while certain hybrid orders can be obtained by condensing excitations in a stack of initially decoupled, two-dimensional topological orders.

\end{abstract}
\maketitle
  \hypersetup{linkcolor=blue}
\tableofcontents
\hypersetup{linkcolor=magenta}

\section{Introduction}
Non-Abelian anyons are exotic fractionalized excitations that arise in two-dimensional quantum systems with long-ranged entanglement, which provide promising avenues for performing fault-tolerant topological quantum computation\cite{NayaketalTQCreview08}. A large class of non-Abelian anyons can arise as the excitations of a gauge theory for a finite, non-Abelian gauge group $G$. Exactly solvable lattice models for these phases have provided insight into the universal properties of the gauge charges and fluxes in these states of matter \cite{Kitaev2003,BombinMartin-Delgado2008_2}, which are in one-to-one correspondence with the irreducible representations (charges) and conjugacy classes (fluxes) of the group $G$. A generalization of these lattice models can be used to study the deconfined phases of $G$ gauge theories in higher spatial dimensions \cite{LevinGu2012,MoradiWen15,Haegemanetal15,Yoshida2017}.

More recently, novel states of matter where the gapped excitations have highly restricted mobility, called fracton orders, have been discovered in three spatial dimensions \cite{Chamon2005,Haah2011,Yoshida2013,VijayHaahFu2015,VijayHaahFu2016}. Fracton orders are similarly obtained by gauging a paramagnet with subsystem symmetry\cite{CobaneraOrtizNussinov2011,VijayHaahFu2016,Williamson2016,KubicaYoshida2018,Pretko2018,ShirleySlagleChen2019,Radicevic2019}, defined as symmetry transformations along extensive sub-regions of the lattice. The X-cube model\cite{VijayHaahFu2016}, for example, can be obtained from simultaneously gauging three intersecting planar symmetries.

\begin{table*}[t!]
    \centering
        \caption{ $(G,N)$ {\bf Gauge Theories:} For a given finite group $G$, the three-foliated $(G,N)$ gauge theory constructed for different choices of normal subgroup $N$ can realize hybrid fracton orders that interpolate between a three-dimensional $G$ topological order and a pure fracton order. The gapped excitations in theses orders above have mobilities which are indicated by the color-coding: a mobile gauge charge (red), a mobile flux loop (blue), fracton (pink) and lineon (green). For a one-foliated $(G,N)$ gauge theory, the hybrid order contains planon excitations rather than lineons or fractons.}
    \begin{tabular}{c}
    \includegraphics[scale=0.5]{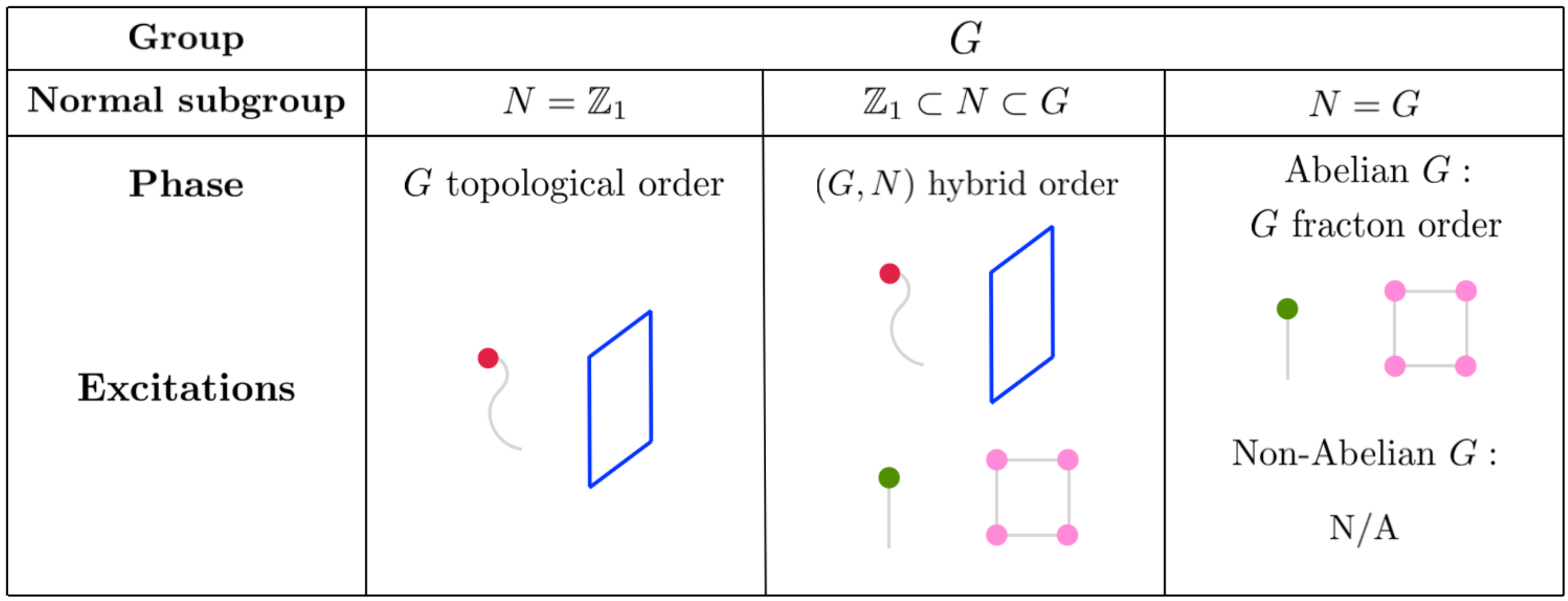}
    \end{tabular}
    \label{tab:3orders}
\end{table*}

A natural question to ask is whether one can similarly obtain novel phases which hosts non-abelian fracton excitations by ``gauging" short-range-entangled systems with non-Abelian subsystem symmetry groups. In this general setting, it is generally not possible to gauge the full non-Abelian subsystem symmetry group; only an Abelian subgroup of the symmetry group can be gauged, thus giving rise to an Abelian fracton order\footnote{Suppose the system of interest has three intersecting planar symmetries, each of which transforms a given plane under a non-abelian group $G$. The group commutator of two intersecting planar symmetries acts only on the intersection line. Therefore, the system has additional line symmetries given by the commutator subgroup $[G,G]$. Repeating the same argument with a third plane, we conclude that the system has a local symmetry $[G,G]$ at each site, and the only physical planar symmetry that can be gauged is the quotient group $G/[G,G]$, which is always abelian.}. Other examples of non-Abelian fracton orders have been recently proposed, by strongly coupling non-abelian topological orders in 2D and 3D  \cite{Vijay2017generalization,PremHuangSongHermele2019,AasenBulmashPremSlagleWilliamson20,Wen20,Wang20,WilliamsonCheng20}, by gauging the abelian subsystem symmetries in certain symmetry-protected topological phases \cite{SongPremHuangMartinDelgado19,StephenGarre-RubioDuaWilliamson2020}, or gauging a global symmetry which permutes fracton excitations in an abelian fracton order\cite{BulmashBarkeshli2019,PremWilliamson2019,WangXu19,WangXuYau19,WangYau19}. Nevertheless, a more general procedure for obtaining non-Abelian fracton orders given a non-Abelian symmetry group -- in direct analogy with the two-dimensional quantum double -- remains elusive. 

In this work, we introduce a general construction to obtain topological quantum orders which host immobile, non-Abelian point-like excitations.  These orders may be understood as the deconfined phase of an exotic gauge theory which is obtained by gauging a paramagnet with global symmetry group $G$ along with subsystem symmetries of an abelian normal subgroup $N$ arranged in a specified geometry. These two symmetries are not independent, since products of the subsystem symmetry transformation across distinct layers can generate a global symmetry transformation, which forms a normal subgroup of the global symmetry group $G$.  We refer to the symmetry of such a paramagnet a \textit{2-subsystem symmetry} denoted $(G,N)$ and the long-range-entangled states that result from gauging such a symmetry\textit{ $(G,N)$ gauge theories}. For a non-Abelian group $G$, there are choices of an Abelian normal subgroup  $N$ such that our construction yields a phase hosting non-Abelian fracton excitations.  

We note that when the specified geometry of the subsystem symmetries in the short-range-entangled state consists of $n$ sheaves of intersecting planes, we will refer to the $(G,N)$ hybrid fracton order obtained by gauging these symmetries as \textit{$n$-foliated}. We emphasize that this terminology only denotes the geometry of the symmetry group before the gauging procedure, and does not immediately imply that two dimensional topological orders can be ``exfoliated" via a finite-depth quantum circuit from these $(G,N)$ gauge theories. The latter definition of ``foliated" fracton orders has been previously used in the literature\cite{ShirleySlagleWangChen2018,ShirleySlagleChen19,ShirleySlagleChen19_2}.

The novelty of our construction is twofold. First, we demonstrate that $(G,N)$ gauge theories simultaneously host both mobile point and loop excitations akin to the quantum double model, and point excitations with restricted mobility, as seen in fracton orders. This is illustrated in Table \ref{tab:3orders}. Because of the non-trivial fusion and braiding of these two types of excitations, the $(G,N)$ gauge theories obtained cannot be in the same phase as the tensor product of a pure 3+1d topological order and a pure fracton order. We note that particular examples of $(G,N)$ gauge theories for Abelian $G$ were explored in Ref. \cite{TJV1}, where it was shown that the resulting orders exhibit a ``hybrid" phenomenology, and host both excitations resembling those of a $G/N$ gauge theory and those of an Abelian fracton order based on subsystem symmetry group $N$.  

In addition, we argue a general relation between the properties of the groups $G$ and $N$ to the features of the fractionalized excitations in the $(G,N)$ gauge theory. A certain set of point-like excitations (``charges") transform as irreducible representations (irreps) of $G$, while another set of excitations (``fluxes") are labeled by conjugacy classes of $G$, reminiscent of the labeling of excitations in the two-dimensional quantum double.  Importantly, a key property we argue is that the irreps that correspond to non-trivial irreps of $Q\cong G/N$ are fully mobile excitations, and otherwise are fully immobile. Similarly, conjugacy classes that correspond to non-trivial conjugacy classes of $Q$ are loop excitations, and otherwise are point excitations with restricted mobility. 

To concretely study these properties, we construct an exactly solvable commuting projector model for any choice of input groups $(G,N)$ and for which the planar subsystem symmetries intersect in a specified geometry. For the case of three intersecting planar symmetries (as in the construction of the X-cube model), our Hamiltonian takes the general form
\begin{align}
    H_{(G,N)} = -\sum_v\bs A^G_v - \sum_{r=x,y,z}\sum_{c} \bs B^N_{c,r} - \sum_p \bs B^Q_p.
    \label{equ:hybridXcubeschematic}
\end{align}
Here, violations of $\bs A^G_v$ for each vertex and point charges, where, depending on the type, can be either fractons or fully mobile. Furthermore, violations of $\bs B^Q_p$ are loop excitations, and those of $\bs B^N_{c,r}$ are lineons, which are mobile in the direction $\hat r$. When the subsystem symmetries of the paramagnet lie along parallel planes (1-foliated), we are also able to derive the operators which create specific excitations in the corresponding $(G,N)$ gauge theory. 

The solvable Hamiltonian includes, as limiting cases, both conventional fracton orders and the three-dimensional quantum double.  Specifically, when the group $N=\ZZ_1$, the model reduces to the $3+1$d quantum double model of $G$ on a cubic lattice.
When the global symmetry group $G$ is Abelian, and $N = G$, the model is equivalent to a fracton order with subsystem symmetry gauge group $N$. It is when $Q$ and $N$ are both non-trivial groups and when $G$ is a non-trivial extension of $Q$ by $N$ that we find the presence of hybrid orders, which provide an interplay between topological order and fracton orders.\footnote{When $G$ is a trivial extension of $Q$ by $N$, the model is a tensor product of a $Q$ quantum double model and a fracton order with subsystem symmetry gauge group $N$.} Such a hybrid model always hosts non-Abelian fractons when $G$ is non-Abelian. That is, there exist irreducible representations of $G$ with dimension greater than one that displays restricted mobility.

    \begin{table}[t!]
\caption{{\bf Three-foliated, hybrid fracton orders corresponding to groups $G=\ZZ_4$, $S_3$, $D_4$, and $Q_8$, and an indicated choice of normal subgroup.} The fracton excitations in $(G,N)$ gauge theories are labeled by irreducible representation (irreps) of $G$. Here,  $\boldsymbol{s}$, and $\boldsymbol{r}$ are different sign representations, while $\boldsymbol{2}$ is a two-dimensional irreducible representation of $G$.  Finally, $\boldsymbol{e}$ is the generating irrep of $\mathbb{Z}_{4}$.  References are included for models that have appeared in previous works. }
        \begin{tabular}{|c|c|c| c|c|}
    \hline
   \multirow{2}{*}{ $G$} &\multirow{2}{*}{ $N$} & \multicolumn{2}{c|}{Fractons} & \multirow{2}{*}{Refs.}\\
   \cline{3-4}
   && Non-Abelian & Abelian &\\
    \hline
     $\ZZ_4$ & $\ZZ_2$ & - & $\boldsymbol{e}$ & \cite{TJV1}\\
     \hline
    $S_3$ & $\ZZ_3$ & $\bs 2$ & - & \cite{TuChang21}\\
    \hline
\multirow{3}{*}{$D_4$} & $\ZZ_2$ &$\bs 2$& - & \cite{AasenBulmashPremSlagleWilliamson20}\\
\cline{2-5}
 & $\ZZ_2^2$ &$\bs 2$&$\bs s$, $\bs{rs}$ & \cite{BulmashBarkeshli2019,PremWilliamson2019,StephenGarre-RubioDuaWilliamson2020,TuChang21}\\
 \cline{2-5}
 & $\ZZ_4$ &$\bs 2$&$\bs r$, $\bs{rs}$& - \\
 \hline
 \multirow{2}{*}{$Q_8$} & $\ZZ_2$ &$\bs 2$& -& - \\
 \cline{2-5}
 & $\ZZ_4$ &$\bs 2$&  $\bs r$, $\bs {rs}$& - \\
 \hline
    \end{tabular}
    \label{tab:example-main}
\end{table}

By systematically studying $(G,N)$ gauge theories, our work provides a unified way of obtaining non-Abelian fracton orders studied in the literature, while also yielding new fracton orders. Many previous non-Abelian fracton orders that have been considered correspond to $(G,N)$ gauge theories, where the non-Abelian group $G$ takes the particular form $G \cong N\rtimes Q$ where $N$ is an Abelian group. Physically, these orders are obtained by starting with an Abelian fracton order (with gauge group $N$) whose excitations are permuted under the action of a global symmetry group $Q$, and then gauging this permutation symmetry. 
The difficulty in these previous examples is that for each $(G,N)$ symmetry pair, one needs to design Abelian fracton models with the desired enriching symmetry. The difficulty is resolved using our construction.

The $(D_4, \ZZ_2^2)$ gauge theory reproduces the model in Refs. \onlinecite{BulmashBarkeshli2019,PremWilliamson2019,StephenGarre-RubioDuaWilliamson2020}, where the symmetry that exchanges a pair of $\ZZ_2$ X-cube models is gauged, or by gauging an SPT with protected by both a $\ZZ_2$ global symmetry and a $\ZZ_2^2$ subsystem symmetry. The $(D_4,\ZZ_2)$ gauge theory produces the same excitations found in the $D_4$ defect network model constructed in Ref. \onlinecite{AasenBulmashPremSlagleWilliamson20}.  We show that the $(D_4, \ZZ_4)$ gauge theory is a third model which hosts $D_4$ non-Abelian fractons, previously conjectured in Ref. \onlinecite{BulmashBarkeshli2019}. We explicitly construct this model and confirm that it can be obtained from gauging the charge conjugation symmetry of the $\ZZ_4$ X-cube model. 

Non-Abelian groups $G$ that arise from a general group extension can also be chosen, yielding a rich array of non-Abelian fracton orders that have not been previously studied. An example we explicitly construct is a model with $Q_8$ (quaternion) non-abelian fractons, which cannot be obtained by gauging a symmetry that only permutes or fractionalizes the fracton excitations\footnote{Mathematically, this follows from the fact that the group extension \begin{equation*}
    \displaystyle 1 \rightarrow \ZZ_4 \rightarrow Q_8 \rightarrow \ZZ_2 \rightarrow 1
\end{equation*} is neither central nor split.}. In Table \ref{tab:example-main}, we list the non-abelian and abelian fractons that appear in the hybrid fracton order for various choices of $G$ and $N$.

A number of open questions remain about the nature of hybrid fracton orders and $(G,N)$ gauge theories.  First, the behavior of the hybrid orders presented here under entanglement renormalization\cite{Haah14,DuaSarkarWillamsonCheng20} is not known. In particular, it is not apparent if the foliation structure of the $N$ subsystem symmetries in a short-range-entangled state implies that two-dimensional topological orders can be exfoliated from the corresponding $(G,N)$ gauge theory using a finite-depth quantum circuit \cite{ShirleySlagleWangChen2018,ShirleySlagleChen19,ShirleySlagleChen19_2}.
 Second, a possibility not considered in this work is that the short-range-entangled states can be put into a symmetry-protected topological order with respect to the $(G,N)$ symmetry before gauging.  The properties of such twisted hybrid fracton orders are outside the scope of this work.  Third, $(G,N)$ gauge theories based on continuous groups remain to be systematically studied.  Fourth, for 1-foliated $(G,N)$ gauge theories with non-Abelian loop excitations, the loops generally exhibit a quantum dimension which grows exponentially in the loop length.  The universal properties of this excitation (e.g. three-loop braiding) remain to be explored, as well as possible uses of these excitations for robust quantum information storage and computation. Finally, it would be interesting to see a general field theory construction for $(G,N)$ gauge theories. One such candidate has been recently proposed in Ref. \onlinecite{HsinSlagle21}.

\subsection{Summary}

We now provide an outline of our paper. The first part of this work studies $(G,N)$ gauge theories which give rise to 1-foliated fracton orders.  In this case, the short-range-entangled state exhibits subsystem symmetries along non-intersecting planes.  The foliation structure of the resulting gauge theory is also apparent from a complementary constructions, by starting with a decoupled stack of two-dimensional $G$ gauge theories, and proliferating an appropriate set of gapped excitations that couple the layers together. 

In Sec. \ref{sec:S3Z3}, we present the detailed properties of a non-Abelian one-foliated  $(G,N)$ gauge theory with $(G,N) = (S_3,\ZZ_3)$ in detail\footnote{The simplest non-trivial abelian model is $(\ZZ_4,\ZZ_2)$ and has been presented in Ref. \onlinecite{TJV1}.}.  In addition to non-abelian point-like excitations which are restricted to move within planes (planons), the resulting order contains a loop-like excitation with a quantum dimension that grows {exponentially} in its length.

 The intricate connection between the choice of groups $G$ and $N$ and the properties of the excitations in the one-foliated gauge theory are clarified by studying the properties of the $(D_4,N)$ gauge theories, where $D_{4}$ (the dihedral group of eight elements) admits three distinct normal subgroups $N$. Each choice leads to different mobility constraints of the excitations.  More generally, for the one-foliated $(G,N)$ gauge theories, we demonstrate a general connection between $G$, $N$ and the properties of the deconfined excitations in Sec. \ref{sec:conditions}.  We show that, like the quantum double model, these gauge theories admit excitations which are labeled by irreducible representations (irreps) of $G$ (charges) and conjugate excitations (fluxes) which are labeled by conjugacy classes of $G$, in analogy with the three-dimensional gauge theories with finite group $G$.  In addition, however, we show that given the projection map from $G$ to the quotient group $Q$,
 \begin{enumerate}
    \item A charge excitation is mobile iff the corresponding irrep of $G$ can be pulled back from a non-trivial irrep in $Q\cong G/N$. Otherwise the charge has restricted mobility.
    \item A flux corresponding to a conjugacy class of $G$ is a loop-like excitation iff the conjugacy class is non-trivial when pushed forward to $Q$.  Otherwise, the flux is point-like, with restricted mobility.
 \end{enumerate}
 A detailed description of these conditions, as well as the conditions for mobile charge and flux loop excitations is provided in Sec. \ref{sec:conditions}.

In Sec. \ref{sec:1foliatedGN}, we present an exactly solvable model for a 1-foliated $(G,N)$ gauge theory, for an arbitrary choice of $G$ and $N$. We also verify the properties of the charges and fluxes explicitly by constructing the operators which create these excitations.

In Sec. \ref{sec:generalGN}, we present a commuting projector model whose ground state realizes any 3-foliated $(G,N)$ gauge theory. We argue that the same relationship between $G$, $N$, and the mobility of the emergent excitations in the 1-foliated $(G,N)$ gauge theory, holds for the 3-foliated case. The 3-foliated gauge theory has no obvious condensation picture, especially because there is no known fracton model with non-abelian gauge group. We demonstrate that the charge excitations are in correspondence with irreps of $G$, and we observe the mobility constraints of these excitations explicitly by studying the lattice Hamiltonian.  

\section{Example: 1-foliated $(S_3,\ZZ_3)$ Gauge Theory}\label{sec:S3Z3}
In this section, we discuss the simplest non-abelian group $G=S_3$, for which the only proper and non-trivial normal subgroup is $N=\ZZ_3$. Our discussion here will follow very closely to that of the 1-foliated $(\ZZ_4,\ZZ_2)$ gauge theory (i.e. the hybrid toric code layers of Ref. \onlinecite{TJV1}).

\subsection{$(S_3,\ZZ_3)$ paramagnet}
We start with an short range entangled (SRE) phase, consisting of $L$ independent copies of a two-dimensional paramagnet with $S_3 = \inner{r,s| r^3=s^2=(sr)^2=1}$ symmetry (for example, the paramagnet of the 3-states Potts model). The full symmetry group of the stacked layers is $S_3^L$. The elementary charged excitations in each layer transform under irreducible representations (irreps) of $S_3$. The irreps are
\begin{enumerate}
    \item  $\bs 1$: the trivial irrep of dimension 1
    \item $\bs s$: the sign irrep of dimension 1
    \item $\bs 2$: the faithful irrep of dimension 2
\end{enumerate}
The non-trivial fusion rules of the charges are given by
\begin{align}
    \bs s \otimes \bs s &=\bs 1,\\
   \bs 2\otimes \bs 2 &= \bs 1 \oplus \bs s
   \label{equ:S3_2fusion}
\end{align}
Next, we will explicitly break the symmetry from $S_3^L$ down so that each layer only has a $\ZZ_3$ symmetry, while retaining a global $S_3$ symmetry. This can be done by adding couplings between adjacent layers that allow only the charges $\bs s$ to tunnel between layers. As a group, we denote the remaining symmetry $(S_3,\ZZ_3)$. The global symmetry -- defined as the product of $S_3$ symmetries in all layers -- is still preserved, and so we can still label the charged excitations as irreps of $S_3$. However, because we no longer have conservation of $\bs s$ in each layer, the subsystem symmetry is broken down to $N=\ZZ_3$ in each plane. Under this subgroup, we note that $\bs s$ transforms trivially, which is consistent with the fact that it is allowed to hop between layers. Furthermore, the 2D irrep $\bs 2$ reduces down to a direct sum of the two non-trivial irreps of $\ZZ_3$ (which we will call $\bs \omega$ and $
\bs {\bar \omega}$). Importantly, since it is still a non-trivial representation under $N$, the $\bs 2$ excitation is a planar excitation. However, Eq. \eqref{equ:S3_2fusion} implies that the outcome of fusing two $\bs 2$ excitations is always mobile.

The result of breaking $S_3^L$ down to $(S_3,\ZZ_3)$ also splits irreps of $S_3^L$ into distinct irreps of $(S_3,\ZZ_3)$. Consider two distinct planes $p_1$ and $p_2$ and consider the bound state of $\bs 2_{p_1}$ and $\bs 2_{p_2}$. After breaking the symmetry, this bound state breaks into two different irreps of $(S_3,\ZZ_3)$ of dimension two, which we will denote as $\bs 2_{p_1,p_2}$ and $\bs 2_{p_1,\bar p_2}$. These non-abelian charge excitations transform as the $\bs 2$ of $S_3$, while under the planar symmetries $\ZZ_3^{p_1}$ and $\ZZ_3^{p_2}$, they transform as
   $ \bs \omega_{p_1}\bs \omega_{p_2} \oplus \bs{\bar \omega}_{p_1} \bs{\bar \omega}_{p_2}$ and $ \bs \omega_{p_1} \bs{\bar \omega}_{p_2} \oplus \bs{\bar \omega}_{p_1} \bs \omega_{p_2}$, respectively. In general, (without the constraint of locality) there are $\frac{3^L-1}{2}$ distinct non-abelian excitations which transform as the $\bs 2$ of $S_3$, but differ in how they are charged under each $\ZZ_3$ planar symmetry.

Let us now gauge the remaining symmetry of the paramagnet. The properties of the charge excitations of the SRE state carry over to the gauge charges of the resulting hybrid order. Qualitatively, we first gauge the $\ZZ_3$ planar symmetries to obtain a stack of $\ZZ_3$ toric codes, where the irreps $\bs \omega$ and $\bs {\bar \omega}$ of the $\ZZ_3$ symmetry are promoted to the anyons $e$ and $\bar e$, respectively in each layer. The global symmetry is now reduced to the quotient group $Q=G/N=\ZZ_2$, which acts as charge conjugation on the toric code anyons as $e_p \leftrightarrow \bar e_p$ and $m_p \leftrightarrow \bar m_p$, where the bar denotes the antiparticle and $p$ denotes each plane. Because of this, the global symmetry enriches the stack of $\ZZ_3$ toric codes. Finally, we can gauge the $Q$ global symmetry. The charges $\bs s$ of $Q$ are promoted to gauge charges of the hybrid model, while the superposition of $e_p$ and $\bar e_p$ in each plane is promoted to the non-abelian particle $\bs 2_p$ of the hybrid model. The remaining charge excitations are also obtained from superposition of $e_p$ in various planes. For example $\bs 2_{p_1,p_2}$ and $\bs 2_{p_1,\bar p_2}$ are obtained from $e_{p_1}e_{p_2} \oplus\bar e_{p_1}\bar e_{p_2} $ and $e_{p_1}\bar e_{p_2} \oplus\bar e_{p_1} e_{p_2} $, respectively.

More generally, a dyon excitation with quantum dimension $2$ can be constructed from a superposition of an arbitrary excitation in the stack of $\ZZ_3$ toric code and it's charge conjugated partner. We will call such general dyon excitation $\bs \alpha$, which can be uniquely labeled by a length $2L$ vector $(\alpha_1,\ldots,\alpha_{2N})$ with entries in $\ZZ_3$ (up to an equivalence of negating all the entries). Each excitation $\bs \alpha$ is then constructed from the superposition
\begin{align}
\left(\prod_{i=1}^L e_{p_i}^{\alpha_i}m_{p_i}^{\alpha_{L+i}} \right )\oplus \left(\prod_{i=1}^L \bar e_{p_i}^{\alpha_i}\bar m_{p_i}^{\alpha_{L+i}} \right)
\end{align}
after gauging the $\ZZ_2$ global symmetry. There are $\frac{3^{2L}-1}{2}$ such non-abelian planon excitations.

\begin{table}[t!]
    \centering
        \caption{{\bf Summary of excitations in stacks of $\mathcal D(S_3)$ and the 1-foliated $(S_3,\ZZ_3)$ gauge theory.} The quantum dimension $d$ for each type of excitation is listed. The phase of $\mathcal D(S_3)$ stacks becomes the phase described by the $(S_3,\ZZ_3)$ gauge theory, after the pair of $\bs s$'s from adjacent layers are condensed. The excitations $\bs \alpha$ are excitations that split from a fusion of non-abelian excitations between different layers in the stack of $\mathcal D(S_3)$. }
    \begin{tabular}{|r l|c|c|c|c|}
    \hline
    \multicolumn{2}{|c|}{\multirow{2}{*}{ Excitation label}} & \multicolumn{2}{c|}{$\mathcal D(S_3)$ stacks}& \multicolumn{2}{c|}{$(S_3,\ZZ_3)$} \\
    \cline{3-6}
    && Type & $d$ & Type & $d$ \\
    \hline
        $(1_1,\bs 1)$ &$\equiv \bs 1$ & mobile & 1 & mobile & 1\\
        $(1_1,\bs s)$ &$\equiv \bs s$ & planon & 1 & mobile & 1\\
        $(1_1,\bs 2)$&$\equiv \bs 2$ & planon & 2 & planon & 2\\ 
        $(1_r,\bs 1)$& $\equiv 1_r$ & planon & 2 & planon & 2\\
        $(1_r,\bs \omega)$& & planon & 2 & planon & 2\\
        $(1_r, \bs{\bar \omega})$& & planon & 2 & planon & 2\\
        $(1_s,\bs 1)$&$\equiv 1_s$ & planon & 3 & loop & $3^{2l}$\\
        $(1_s,\bs s)$& & planon & 3 & loop & $3^{2l}$\\
        $\bs \alpha$ & & - & - & planon & 2\\
        \hline
    \end{tabular}
    \label{tab:S31foliated}
\end{table}

To complete our analysis, we need to study the flux loop excitation, which corresponds to a defect of the global symmetry $Q$ before gauging. However, we find it more enlightening to consider an alternative construction of this hybrid model via a condensation transition. First, we temporarily neglect the interlayer couplings and gauge the full $S_3^L$ symmetry of the stacked 2D paramagnets. The result is a stack of $L$ quantum double models $\mathcal D(S_3)$. There are eight types of anyons in each layer, which can be labeled by a conjugacy class and an irrep of the corresponding centralizer, as summarized in Table \ref{tab:S31foliated}. Note that when either the conjugacy class or irrep is trivial, we will often refer to the excitation using its other non-trivial label. Such excitations, are the pure charges and pure fluxes of the model, respectively.

To recover the hybrid model, we can recover the interlayer hoppings by inducing a condensation transition between a pair of $\bs s$ anyons in every adjacent layer. As a consequence, the $\bs s$ in each layer are now in the same superselection sector in the condensate phase, making the $\bs s$ particle mobile. The full set of planon excitations can be derived by considering the splitting of products of the $\bs 2$ planon in each layer.

Now, the pure flux $1_s$ braids non-trivially with $\bs s$, and is therefore confined. However, a composite loop excitation, composed of a pair of $1_s$ excitations in each layers braids trivially with the condensate, and therefore remains deconfined. We will use the same label of the original anyon to label this loop excitation in the hybrid model. That is, we will call this the ``$1_s$ loop".

Before investigating the properties of the loop let us first argue that this loop excitation does not fuse into anything that forms the condensate, and therefore does not split. First, consider the following fusion rule for two $1_s$ anyons
\begin{align}
    1_s \otimes 1_s = \bs 1 \oplus \bs 2 \oplus 1_r \oplus  (1_r,\bs \omega) \oplus (1_r, \bs{\bar \omega}).
    \label{equ:1sfusion}
\end{align}
We see that the fusion outcomes do not involve the charge $\bs s$. Similarly, the fusion of $1_s$ with all the other anyons except $\bs s$ and $(1_s,\bs s)$ (the bound state of the two) does not involve $\bs s$. This guarantees that the fusion $1_s$ loop with some other excitation not involving $\bs s$ will have at most one trivial superselection sector as a fusion outcome. Thus, after condensing pairs of $\bs s$ in adjacent layers, the fusion of the $1_s$ loop with the remaining deconfined particles will have a unique trivial superselection sector. This guarantees that the $1_s$ loop does not split.

We can now discuss the peculiar properties of the $1_s$ flux loop. First, since the $1_s$ anyon before the condensation has quantum dimension $3$, the resulting flux loop has quantum dimension $3^{2l}$, where $l$ is the number of layers in which the flux loop pierces\footnote{ In general, the flux loop can pierce a layer multiple times, in which case we must count all the number of pierces.}. This extensive quantum dimension can be seen explicitly by fusing two identical flux loops. Following Eq. \eqref{equ:1sfusion}, the fusion of two identical $1_s$ loops in the hybrid order will result in a product of the given fusion outcomes at the $2l$ points where the loop pierces a 2d layer (with further splitting into the planons $\bs l$ if two non-abelian planons in different layers are fused together).

Let us remark that ``shape dependence'' of the quantum dimension of such flux loops was pointed out in Refs. \onlinecite{BulmashBarkeshli2019,PremWilliamson2019} in a model with three foliations. For 1-foliated hybrid models, this curious property is demystified from the condensation construction. In fact, one can see what happens to the quantum dimension when we shrink the loop excitation so that the number of layers pierced decreases by one. In the layer where the loop no longer pierces, shrinking the loop away is just equivalent to fusing two $1_s$ anyons away before the condensation, and therefore in that layer, we will be left with the planons $ \bs 1 \oplus \bs 2 \oplus 1_r \oplus  (1_r,\bs \omega) \oplus (1_r, \bs{\bar \omega})$. This decreases the quantum dimension of the loop by $3^2$.

Lastly, it is perhaps worth pointing out the nature of the loop excitation $(1_s,\bs s)$ in the hybrid model. This is simply the bound state of the mobile particle $\bs s$ and the loop $1_s$. Indeed, the fusion rules are identical to the anyons of $\mathcal D(S_3)$:
\begin{align}
    1_s \otimes \bs s = (1_s,\bs s).
\end{align}
This bound state is well-defined because both excitations are fully mobile.

\begin{figure}[t]
    \centering
    \includegraphics[scale=0.25]{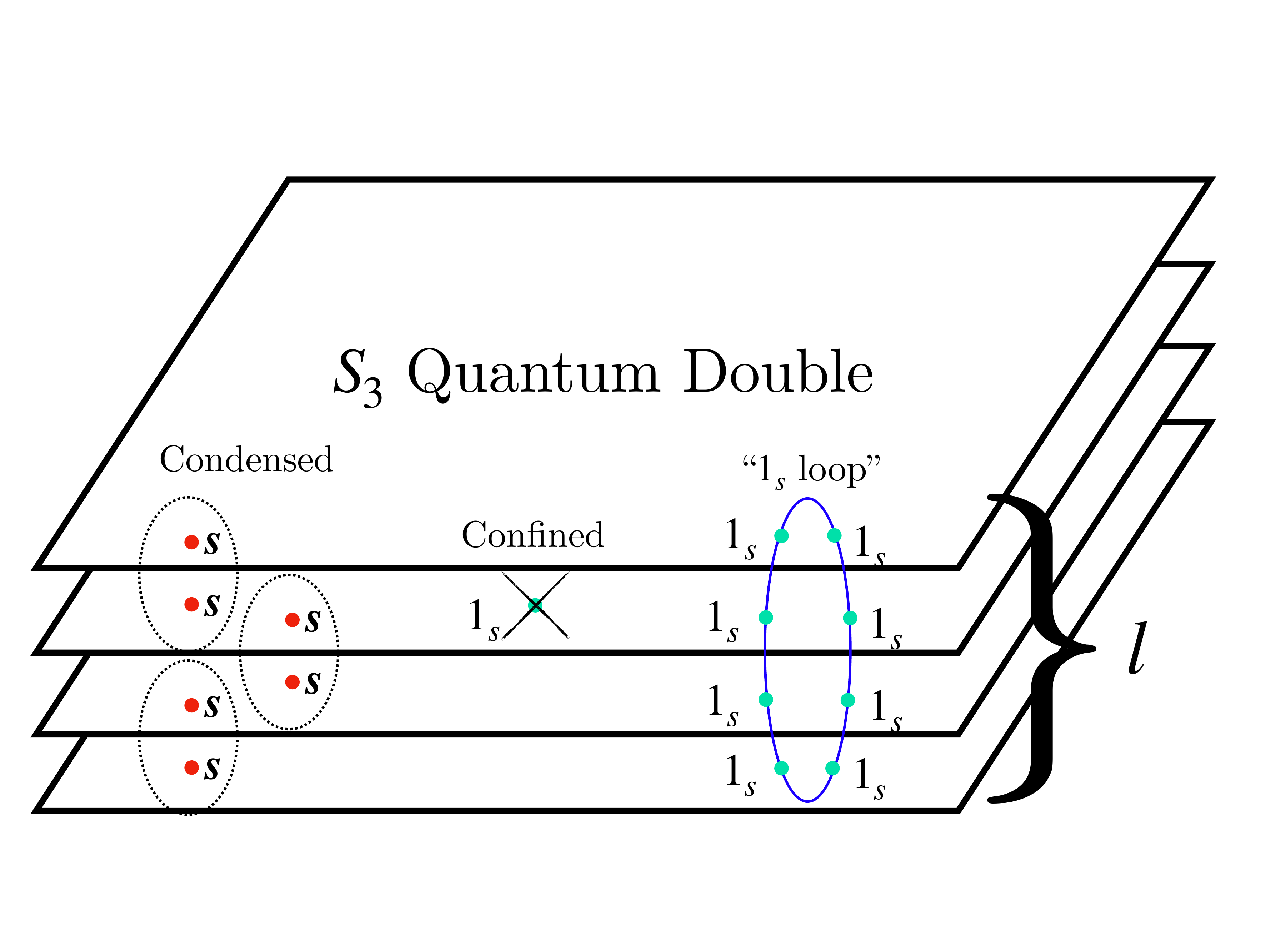}
    \caption{Alternative construction of the 1-foliated ($S_3,\ZZ_3$) hybrid model via a condensation of $\bs s$ pairs in adjacent $\mathcal D(S_3)$ layers. As a result, the flux $1_s$ is confined, but a flux loop composed of $1_s$ pairs in $l$ layers survives the transition. The quantum dimension of this loop excitation is $3^{2l}$.}
    \label{fig:S31foliated}
\end{figure}

\section{Conditions for mobility}\label{sec:conditions}
In this section, we will state the conditions for which a charge of the hybrid fracton model will have restricted mobility in terms of the input data $G$ and $N$. First, we give examples of a 1-foliated $(D_4,N)$ Gauge Theory in Sec. \ref{sec:D4examples}. In contrast to the $(\ZZ_4,\ZZ_2)$ model in Ref. \onlinecite{TJV1} and $(S_3,\ZZ_3)$ in the previous section, there are multiple non-trivial normal subgroups for $D_4$. We will point out how this choice is reflected in the mobility of the particles, which in turns uniquely determines the type of anyons we have to condense in the layer construction. Then, in Sec. \ref{sec:Mobilitygeneralgroup} we give a precise statement of the mobility constraints for a general choice of $G$ and $N$.

\subsection{1-foliated $(D_4,N)$ Gauge Theory}\label{sec:D4examples}

We consider the dihedral group $G=D_4 \equiv \langle r,s| r^4 =s^2=(sr)^2=1\rangle$. The quantum double model $\mathcal D(D_4)$ consists of 22 anyons. For simplicity, we will only discuss the pure charge and flux excitations. There  are five pure charges given by the different irreps of $D_4$: the trivial irrep $\bs 1$, the 2D faithful irrep $\bs 2$, and three sign reps
\begin{enumerate}
\item $\bs s$ with kernel $\ZZ_4= \inner{r}$
\item $\bs r$ with kernel $\ZZ_2^2= \inner{r^2,rs}$
\item $\bs {rs}$ with kernel $\ZZ_2^2= \inner{r^2,s}$
\end{enumerate}
The sign reps have an abelian $\ZZ_2^2 = D_4/\ZZ_2$ fusion rules and are labeled by a representative element under the quotient group. They can be absorbed into $\bs 2$ by fusion. Lastly, the fusion of two $\bs 2$ irreps is given by
\begin{align}
    \bs 2 \otimes \bs 2 = \bs 1 \oplus \bs s \oplus \bs r \oplus \bs {rs}
\end{align}
The fluxes are labeled by conjugacy classes of $D_4$, which we label by representative elements as $1_1, 1_{r^2}, 1_r, 1_s, 1_{rs}$. They satisfy more involved fusion rules, so we will just note that $1_r, 1_s,$ and $1_{rs}$ are non-abelian fluxes of quantum dimension two, and $1_{r^2}$ is an abelian flux of order two.

We will now demonstrate for the case of $(D_4,N)$ that the resulting hybrid model depends on the choice of $N$. Note that in contrast, for a 2D model (i.e. the single layer limit), the resulting topological order $\mathcal D(D_4)$ -- obtained by first gauging $N$ followed by gauging the quotient group $Q=G/N$ -- does not depend on the choice of $N$.

We start with $L$ layers of a paramagnet each invariant under $D_4$. Again, there are two ways to construct the hybrid model. The first is to add charge-hopping terms to reduce the symmetry from $D_4^L$ down to $(D_4,N)$. One can then gauge the remaining symmetry. Alternatively, one can first gauge the $D_4$ symmetry in each layer to obtain a stack of $D_4$ quantum double models. Then, condense pairs of appropriate gauge charges that we want to make mobile. Let us see what the resulting type of excitations we get for each choice of $N$. 

\begin{enumerate}
\item $N=\ZZ_1, Q=D_4$. The resulting model in this case is the usual $D_4$ gauge theory in 3D, since the subsystem symmetry $N$ is trivial. All charges are mobile. Therefore, in the layer construction, we can achieve this by condensing pairs of all gauge charges in adjacent layers.

\item $N=\ZZ_2, Q=\ZZ_2^2$. We begin with the $(D_4,N)$ paramagnet and first gauge the $N$ planar symmetries. This results in layers of $\ZZ_2$ toric codes enriched by a global $Q$ symmetry. In particular, $Q$ acts projectively on the charge $e$, but acts trivially on $m$. 

We next gauge $Q$ to obtain the $(D_4, \ZZ_2)$ gauge theory. Since $e$ transforms projectively, it is promoted to a particle of quantum dimension 2, in each layer, which we can identify as $\bs 2$. Moreover, the mobile charges under $Q$ are promoted to mobile gauge charges, which are the irreps $\bs r$, $\bs s$, $\bs{rs}$. Therefore, we see that this hybrid model can be obtained in the layered construction by condensing pairs of all sign reps in adjacent layers of $\mathcal D(D_4)$. 

In general, the planons $\bs \alpha$ can be labeled by a length $2L$ vector with entries in $\ZZ_2$. The nature of the planon after gauging depends on whether it transforms linearly or projectively under the symmetry. In particular, the representation is linear when $\sum_{i=1}^L \alpha_{i}$ is even, which will result in an abelian excitation. Otherwise, it will result in a non-abelian excitation with quantum dimension 2. Therefore, there are $2^{2L}-1$ abelian planons and $2^{2L}$ non-abelian planons.

To complete the discussion, we discuss the nature of the fluxes. Since $Q$ does not act projectively on $m$, it is not affected by the gauging, and remains an abelian planon of order two. Thus, we can identify it with $1_{r^2}$. The remaining flux loops, which are promoted symmetry defects of $Q$ must therefore be labeled by $1_{r}$, $1_{s}$, $1_{rs}$ and are non-abelian. In the layered construction, the loop excitations are respectively constructed by a product of the anyons $1_{r}$, $1_{s}$, and $1_{rs}$ of the original $\mathcal D(D_4)$ in each layer, which remains deconfined after the condensation of the three sign rep pairs. All flux loops have quantum dimension $2^{2l}$.

\item $N=\ZZ_4, Q=\ZZ_2$. Gauging the $N$ planar symmetry in the paramagnet results in a stack of $\ZZ_4$ toric codes. The remaining global symmetry is $Q=\ZZ_2$, which acts as ``charge conjugation" by permuting the anyons $e \leftrightarrow \bar e$ and $m \leftrightarrow \bar m$ in each layer. Gauging $Q$ promotes of $e$ and $\bar e$ into a single non-abelian planon $\bs 2$, while $e^2$ is promoted to planon $\bs r$. The mobile charge of $Q$ corresponds to a mobile charge $\bs s$, while $\bs {rs}$ is a bound state of $\bs r$ and $\bs s$, and is therefore a planon. Therefore, this model can be obtained in the layered construction by condensing pairs of $\bs s$ in adjacent layers.

A general planon excitation can be labeled by a length $2L$ vector with entries in $\ZZ_4$ with an equivalence of negating all the entries. The excitation is abelian if all the entries contain only $0$ or $2$. There are $2^{2L}$-1 such planons. In all other cases, it is a non-abelian excitation with quantum dimension $2$.

To see the mobility of the fluxes, we note that the $\ZZ_2$ symmetry does not permute $m^2$, so it remains an abelian planon, which we must identify with the flux $1_{r^2}$. On the other hand, since $Q$ exchanges $m$ and $\bar m$, their superposition gets promoted to a non-abelian planon, labeled by $1_r$. The $1_s$ and $1_{rs}$ anyons in the original $\mathcal D(D_4)$ model are confined, and only a string of them are deconfined, which forms loop excitation of the same label.

\item $N=\ZZ_2^2$. Without loss of generality, we choose $N=\inner{r^2,rs}$.\footnote{The case $N=\inner{r^2,s}$ gives an identical model because of the automorphism $s \leftrightarrow rs$, which swaps irreps $\bs r \leftrightarrow \bs{rs}$ and conjugacy classes $1_{s} \leftrightarrow 1_{rs}$.} Gauging the $\ZZ_2^2$ planar symmetry in the paramagnet results in bilayers of $\ZZ_2$ toric codes where the twist defects $1_{rs}$ and $1_{r^2}$ become the anyons $m_1$ and $m_2$. The remaining global symmetry is $Q=\ZZ_2$, which permutes the anyons $e_1 \leftrightarrow e_2$ and $m_1 \leftrightarrow m_2$. Gauging $Q$ promotes the ``superposition'' of $e_1$ and $e_2$ into the planon $\bs 2$, while $e_1e_2$ is promoted to the planon $\bs{rs}$. The mobile charge of $Q$ gauges to $\bs{r}$, and its bound state with $\bs{rs}$ is another planon $\bs s$. Therefore, we see that model can be obtained in the layered construction by condensing pairs of $\bs r$ in adjacent layers.

A general planon excitation can be labeled by a length $2L$ vector with entries in $\ZZ_2^2$. The excitation is abelian if all the entries are $(0,0)$ or $(1,1) \in \ZZ_2^2 $ In all other cases, it is a non-abelian excitation with quantum dimension $2$.

The global $Q$ symmetry acts trivially on $m_1m_2$, so it remains a planon $1_{r^2}$. The superposition of $m_1$ and $m_2$ becomes a non-abelian planon $1_{rs}$. Finally, $1_{s}$ and $1_r$ are loop excitations, which can be seen as a product of $1_s$ and $1_r$ anyons in each layer, which braids trivially with pairs of $\bs r$ in the layer construction.
\end{enumerate}

A summary of the mobility of the pure charges and fluxes for different choices of $N$ is given in Table \ref{tab:D41foliated}.

\begin{table}[t]
\caption{{\bf Summary of pure charges and fluxes for the 1-foliated model $G=D_4$ with different choices of $N$.} Here $\cdot$ = mobile point, $p$ = planon, and $\emptycirc$ = mobile loop. The general planon excitations which are supported on multiple layers are not included here. Note that $\bs{2}$, $1_{s}$, $1_{r}$, $1_{rs}$ are non-abelian excitations, with possibly restricted mobility, depending on $N$. The two $N=\ZZ_2^2$ differ by a choice of the embedding $\iota:N\rightarrow G$ in the group extension \eqref{equ:exact}, but the resulting models are equivalent up to relabeling the excitations. In the 3-foliated model, the charges and fluxes with restricted mobility are instead fractons and lineons, respectively.}
    \centering
    \begin{tabular}{|c|c|c|c|c|c|c|}
    \hline
    \multirow{2}{*}{excitation} & \multicolumn{5}{c|}{$N$}   \\
    \cline{2-6}
    &$\ZZ_1$&$\ZZ_2$&$\ZZ_4$& $\ZZ_2^2$&$\ZZ_2^2$\\
    \hline
$\bs s$       & $\cdot$       & $\cdot$   & $\cdot$ & $p$ &$p$\\
$\bs r$      & $\cdot$       & $\cdot$   & $p$ & $\cdot$&$p$\\
$\bs{rs}$      & $\cdot$       & $\cdot$   & $p$ & $p$&$\cdot$\\
$\bs 2$       & $\cdot$       & $p$   & $p$ & $p$&$p$\\
\hline
$1_{s}$     & $\emptycirc$      & $\emptycirc$   & $\emptycirc$ & $\emptycirc$ &$p$\\
$1_{r}$     & $\emptycirc$      & $\emptycirc$   & $p$ & $\emptycirc$ & $\emptycirc$\\
$1_{rs}$    & $\emptycirc$      & $\emptycirc$   & $\emptycirc$ & $p$& $\emptycirc$\\
$1_{r^2}$   & $\emptycirc$      & $p$  & $p$ & $p$ & $p$\\
        \hline
    \end{tabular}
    \label{tab:D41foliated}
\end{table}

\subsection{Mobilities of excitations}\label{sec:Mobilitygeneralgroup}

We now consider a hybrid fracton model corresponding to choice $(G,N)$ and an arbitrary geometry of the subsystem symmetries. We will make the assumption that pure charges of this model (that are supported in a single layer per foliation direction) can be labeled by irreps of $G$, and that the pure fluxes can be labeled by conjugacy classes of $G$. (We show in Sec. \ref{sec:1foliatedGN} that this holds for the 1-foliated case by explicitly constructing the excitations in an exactly solvable model. For the 3-foliated case, we can only rigorously show the former, and for now we conjecture the latter to be true)

Accepting such assumption, we are able to show the following. Given the group extension
\begin{equation}
1 \rightarrow N  \xrightarrow[]{\iota} G  \xrightarrow[]{\pi} Q \rightarrow 1
\label{equ:exact}
\end{equation}
\begin{enumerate}
    \item Irreps of $G$ that can be pulled back\footnote{Given a group homomorphism $G  \xrightarrow[]{\pi} Q$, the pullback of a representation $U^q$ is given by $U^{\pi(g)}$. Furthermore, since $\pi$ is surjective, an irrep of $Q$ pulls back to a unique irrep of $G$.} from $Q$ correspond to mobile charges
    \item Irreps of $G$ that are non-trivial\footnote{but not necessarily irreducible} when pulled back to $N$ correspond to charges with restricted mobility.
    \item The conjugacy classes of $G$ whose union forms $\iota(N)$ correspond to fluxes with restricted mobility.
    \item The preimage of conjugacy classes of $Q$ under $\pi$ correspond to flux loops.
\end{enumerate}

First, to argue Claim 2, recall that the nature of the restricted mobility of fractons is due to a conservation law in the fracton model. This conservation law arises from gauging subsystem symmetries, which disallows a single charge to move outside the support on which the subsystem symmetry acts\cite{ShirleySlagleChen2019}. Similarly, we see that any irrep of $G$ that transforms non-trivially under (pulling back to) $N$ will be charged under the $N$ subsystem symmetries of the model, and after the duality, must result in a charge excitation with restricted mobility.

Claim 1 is cleanest to see after gauging the $N$ planar symmetries, where only $Q$ remains as the global symmetry. It follows that any charge under $Q$ must be mobile since there is no extra conservation law preventing $Q$ charges to move between planes.

For completeness, we must show that the two conditions above do not overlap. That is, for each irrep of $G$, exclusively only either claim 1 or claim 2 applies. To show this, we use the fact that the exact sequence \eqref{equ:exact} induces a dual short exact sequence
\begin{align}
   0 \leftarrow \hat N  \xleftarrow[]{\hat \iota} \hat G  \xleftarrow[]{\hat \pi} \hat Q \leftarrow 0 
\end{align}
where $\hat G \equiv \text{Hom}(G, \CC)$ is the Pontryagin dual consisting of characters of $G$. Since $\ker \hat \iota = \text{im }\ \hat \pi$, any irreducible character of $G$ is trivial when pushed via $\hat \iota$ to a character of $N$ iff it can be pushed from a character of $Q$ via $\hat \pi$.

Moving on to the fluxes, in Claim 3, the conjugacy classes whose union forms $N$ correspond to the subsystem symmetry defects of $N$. Upon gauging $N$, they become fluxes with restricted mobility. We can also see which of these fluxes are non-abelian. Conjugacy classes that contain elements from multiple conjugacy classes of $N$ are those that become non-abelian after gauging $Q$. We remark that if the extension is central, then all such fluxes will be abelian.

Finally, for Claim 4, we see that the preimage of conjugacy classes of $Q$ correspond to symmetry defects that are not promoted to flux excitations after gauging $N$. Therefore, they are defects of a global symmetry $Q$ and are therefore loop-like.

Similarly to the charges, for each conjugacy class only one of Claim 3 or Claim 4 will apply because of the exactness of the group extension \eqref{equ:exact}. That is, any flux excitation can be uniquely labeled as either be a loop excitation or a point excitation with restricted mobility.

To give concrete examples, we demonstrate that the $(D_4,N)$ models we considered Sec. \ref{sec:D4examples} are consistent with our claims. For brevity, we use the adjective ``restricted" to refer to an excitation with restricted mobility.

\begin{enumerate}
    \item $N=\ZZ_1$. Since $N$ is trivial, $G=Q$ amd therefore irreps of $G$ can be pulled back from $Q$, and are therefore all mobile. Similarly, all conjugacy classes of $G$ can be pushed forward to $Q$, and are hence loop excitations.
    \item $N=\ZZ_2$. The three sign reps correspond to charges of $Q$, and are therefore all mobile, while $\bs 2$ pulls back to the sign rep of $N$, and is therefore restricted. For fluxes, the non-trivial conjugacy class of $N$ pushes forward to the conjugacy class $1_{r^2}$ of $D_4$, so only this flux has restricted mobility (and is abelian). The remaining non-trivial conjugacy classes are preimages of conjugacy classes in $Q$ and so are loop excitations. We find that the mobility conditions of this gauge group for the 3-foliated case is consistent with the model with $D_4$ non-abelian fractons presented in Ref. \onlinecite{AasenBulmashPremSlagleWilliamson20} using the defect network construction.
    \item $N=\ZZ_4$. The irrep $\bs 2$ pulls back to a 2D reducible representation $\bs \omega \oplus \bs {\bar \omega}$ of $N$, while $\bs r$, and $\bs{rs}$ pull back to the irrep $\bs \omega^2$. Therefore, the irreps $\bs 2$, $\bs r$, and $\bs{rs}$ all have restricted mobility. On the other hand, $\bs{s}$ can be pulled back from the sign rep of $Q$ and is therefore mobile. For fluxes, the conjugacy class $1_{r^2}$ in $N$ maps to $1_{r^2}$ in $G$, while $1_r$ and $1_{r^3}$ in $N$ maps to a single conjugacy class $1_r$ in $G$. Therefore, $1_{r^2}$ and $1_r$ are restricted, the latter being non-abelian. The conjugacy classes $1_s$ and $1_{rs}$ map to the non-trivial conjugacy class in $Q$, so such fluxes are mobile loops.
    \item $N=\ZZ_2^2$. The irreps $\bs 2$, $\bs s$, and $\bs {rs}$ are restricted, since they pull back to charges $(-1,-1)$ , $(1,-1)$, and $(1,-1)$ of $N=\ZZ_2^2$ while $\bs r$ can be pulled back from the sign rep of $Q$, so it is mobile. For fluxes, there are three non-trivial conjugacy classes of $N$, one of which maps to $1_{r^2}$ and two of which maps to $1_{rs}$. Therefore, these conjugacy classes are fluxes with restricted mobility, the latter being non-abelian. The conjugacy classes $1_s$ and $1_r$ map to the non-trivial conjugacy class in $Q$, so such fluxes are mobile loops. The mobility conditions of this gauge group for the 3-foliated case is consistent with the $D_4$ model obtained by gauging the \textsc{swap} symmetry of two X-cube models\cite{BulmashBarkeshli2019,PremWilliamson2019}.
\end{enumerate}

\section{1-foliated $(G,N)$ gauge theory}\label{sec:1foliatedGN}

\begin{figure}
    \centering
    \includegraphics[scale=0.25]{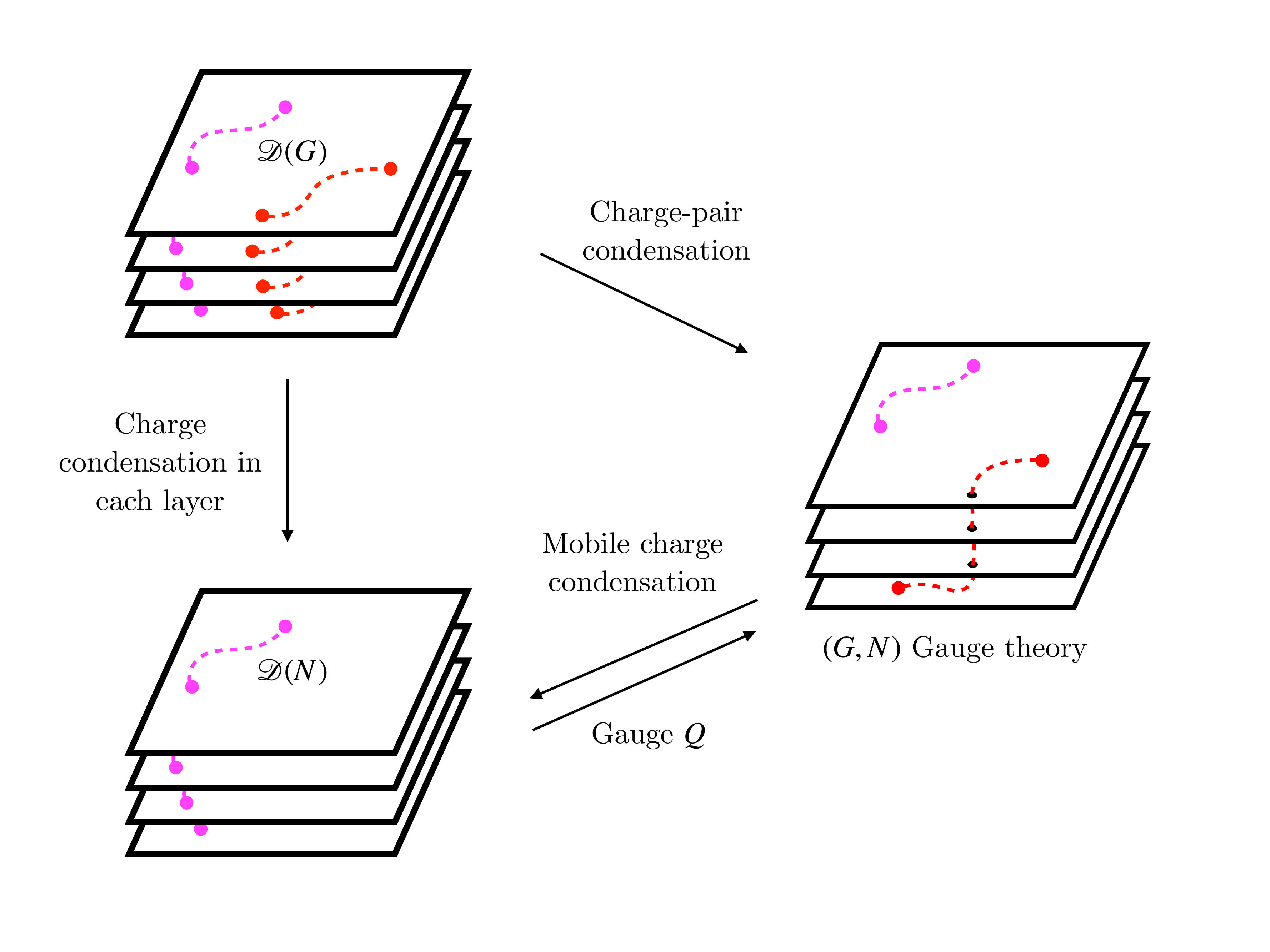}
    \caption{The 1-foliated $(G,N)$ gauge theory can be obtained from starting with a stack of $\mathcal D(N)$ models, and gauging a global $Q$ symmetry enriching the theory. Alternatively, it can also be obtained by starting with a stack of $\mathcal D(G)$ models, and condensing pairs of charges (shown in red) in adjacent layers. The charge pairs that are condensed depend on the choice of $N$, and are exactly the charges that become mobile in the hybrid model. Fluxes are not drawn for simplicity.}
    \label{fig:GNcondensation}
\end{figure}

In this section, we present an exactly solvable model for the 1-foliated $(G,N)$ gauge theory, which can be thought of as a hybrid between a 3D quantum double model and a stack of 2D quantum double models\cite{Kitaev2003,MoradiWen15}. The case where $(G,N)=(\ZZ_4,\ZZ_2)$ is the hybrid toric code layers presented in Ref. \onlinecite{TJV1}. The relation between the $(G,N)$ gauge theory and stacks of familiar $2+1$d topological phases is shown in Figure \ref{fig:GNcondensation}.

We note that, because the subsystem symmetry planes do not intersect, one can in fact choose $N$ to be a non-abelian group. Therefore, the construction in this section holds generally for any finite group $G$.

Given the exact sequence Eq. \eqref{equ:exact} corresponding to the group extension, we place $\mathbb C[G]$ on each edge oriented in the $x$ or $y$ directions of the cubic lattice, equipped with left and right multiplication operators
\begin{align}
L_e^g \ket{g_e} &= \ket{gg_e}, \\
R_e^g \ket{g_e} &= \ket{g_e\bar g},
\end{align}
where $\bar g \equiv g^{-1}$. Furthermore, on each edge oriented in the $z$ direction, we place $\mathbb C[Q]$ with operators
\begin{align}
L_e^q \ket{q_e} &= \ket{qq_e}, \\
R_e^q \ket{q_e} &= \ket{q_e\bar q} ,
\end{align}
Furthermore, for notational convenience, we define
\begin{align}
    q_e' = \begin{cases}
    \pi(g_e); & \text{if }e \text{ is in the $x$ or $y$ directions}\\
    q_e; & \text{if} e \text{ is in the $z$ direction}
    \end{cases}
\end{align}
The Hamiltonian is given by
\begin{align}
    H_{(G,N)} &= -\sum_v \bs A^G_v - \sum_{p_\perp}  \bs B_{p_\perp}\nonumber -  \sum_{p_\parallel} \bs B_{p_\parallel},\\
    \bs A^G_v & = \frac{1}{|G|} \sum_{g \in G} \bs A^g,
    \label{equ:1foliatedGNHam}
\end{align}
where $p_\parallel$ denotes plaquettes in $xy$ plane, and $p_\perp$ denotes plaquettes in the $xz$ and $yz$ planes.

The explicit form of the operators are
\begin{align}
    \bs A_v^g &= \raisebox{-0.5\height}{\includegraphics[scale=1]{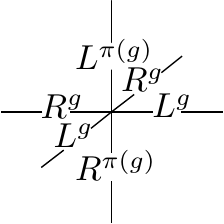}}\\
    \bs B_{p_\perp} &= \delta_{1,q_1'q_2'\bar q_3' \bar q_4'} \Ket{\raisebox{-0.5\height}{\includegraphics[scale=1]{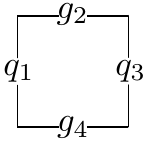}}}\Bra{\raisebox{-0.5\height}{\includegraphics[scale=1]{Bpyz1foliatedGN.pdf}}}\\
    \bs B_{p_\parallel} &= \delta_{1,g_1g_2\bar g_3 \bar g_4} \Ket{\raisebox{-0.5\height}{\includegraphics[scale=1]{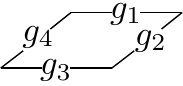}}}\Bra{\raisebox{-0.5\height}{\includegraphics[scale=1]{Bpxy1foliatedGN.pdf}}}
\end{align}

\subsubsection{Excitations and fusion}
As in the quantum double models, excitations of the hybrid model can be obtained by applying appropriate ribbon operators \cite{Kitaev2003,BombinMartin-Delgado2008}. For simplicity, we only discuss how to create pure charge and pure flux excitations.

Charge excitations are labeled by irreps $\bs \mu$ of the $G$, and can be excited at the end points of a direct string $L$. First, let us assume the string is purely in a fixed $xy$ plane denoted $L_{xy}$, consisting of an ordered set of edges $e_1, e_2, \ldots, e_n$. Let $\Gamma^{\bs \mu,G}$, be the matrix representation of the irrep $\mu$ of $G$, then for each $i,i' = 1 \ldots d_{\bs \mu}$, where $d_{\bs \mu}$ is the dimension of the irrep $\bs \mu$, one can construct the following string operator
\begin{align}
    F^{\bs \mu, (i ,i')}_{L_{xy}}=\sum_{\{ g_e\}} \Gamma^{\bs \mu,G}_{i,i'}\left (g_{e_1}^{O_{e_1}}g_{e_2}^{O_{e_2}} \cdots g_{e_n}^{O_{e_n}}\right )  \ket{\{ g_e \}}\bra{\{ g_e \}}
    \label{equ:1foliatedGNstring}
\end{align}
where $\{ g_e \}$ are group elements along  $L_{xy}$ and $O_e$ inverts the group element if the path of the string $L_{xy}$ goes against the orientation of that edge. This ribbon violates $\bs A_v$ only at the end points. Furthermore, under the operators $\bs A^g_v$, the left (right) end point transforms under the representation $\bs \mu$ ($\bar{\bs \mu}$) of $G$.

The above string operator seems to imply that all charge excitations are planons. However, that is not the case. As we have argued in Sec. \ref{sec:Mobilitygeneralgroup}, a charge corresponding to the irrep $\bs \mu$ is actually mobile if $\bs \mu$ can be pulled back from some irrep $\bs \nu$ of $Q$. That is,
\begin{align}
   \Gamma^{\bs \mu,G} (g) = \Gamma^{\bs \nu,Q} (\pi(g)) \ \ \ \ \ \forall g \in G
   \label{equ:pullbackcharacter}
\end{align} 
This can be seen explicitly by constructing the corresponding string operator. For a direct string $L$ (not necessarily confined to a single $xy$ plane), consider 
\begin{align}
    F^{\bs \mu, (i ,i')}_{L}=\sum_{\{ g_e,q_e\}}& \Gamma^{\bs \nu,Q}_{i,i'}\left ({q'}_{e_1}^{O_{e_1}}{q'}_{e_2}^{O_{e_2}} \cdots {q'}_{e_n}^{O_{e_n}}\right ) \nonumber\\
    &\ket{\{ g_e,q_e \}}\bra{\{ g_e,q_e \}}.
    \label{equ:1foliatedGNmobilestring}
\end{align}
This operator also violates $\bs A_v$ only at the end points of the string, and the left (right) end point transform as  $\bs \mu$ ($\bar{\bs \mu}$) of $G$ under $\bs A^g_v$. It is also apparent that by projecting $\bs A^g_v$ to $Q$, the end points transform under the irreps $\bs \nu$ ($\bar{\bs \nu})$ of $Q$. Finally, we remark that if $L$ is in a fixed $xy$ plane, the general string operator in \eqref{equ:1foliatedGNmobilestring} reduces to Eq. \eqref{equ:1foliatedGNstring} defined in a single plane, by applying the definition of the pullback \eqref{equ:pullbackcharacter}.

To complete the argument, we must show that the remaining irreps of $G$ (i.e. the irreps $\bs \mu$ for which Eq. \eqref{equ:pullbackcharacter} does not hold), once created, cannot move out of their respective planes. Intuitively, this is because such a charge, when moved outside its original plane, will violate the conservation law of the new plane it enters. Suppose that such an irrep $\bs \mu$ were mobile. This means that we are able to move a single such charge freely into a new layer. Consider $\bs A^g_{v}$ where $g$ is restricted to the subgroup $N$. Since $\pi(g)=1$, $\bs A^g_{v}$ only acts on each $xy$ plane, and therefore we see that $\prod_\text{plane} \bs A^g_v=1$. However, the charge moved to this new layer must transform non-trivially under $\prod_\text{plane} \bs A^g_{v}$ and is therefore a contradiction.

To conclude, the charges fuse according to the fusion category Rep$(G)$, just as the quantum double model of $G$. The additional structure (reflecting the choice of $N$ in the $(G,N)$ gauge theory) is that certain charges are planons. As a result, we note that generally, such planons can fuse into a direct sum of other charges, some of which are mobile.

Now, we discuss a basis of which one can create general excitations of the hybrid model. Consider the following ribbon operator (identical to those in the quantum double model) which are labeled by two group elements $g$ and $l$.
\begin{align}
    &F^{g,l} \Ket{\raisebox{-0.5\height}{\includegraphics[scale=1]{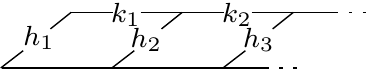}}} \\
    &= \delta_{l,k_1k_2 \cdots} \nonumber \Ket{\raisebox{-0.5\height}{\includegraphics[scale=1]{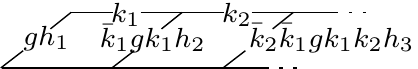}}}
\end{align}
For $g=1$, an appropriate superposition of $F^{1,l}$ (namely the Fourier transform) gives the string operator for the charge excitations Eq. \eqref{equ:1foliatedGNstring}.

Let us now consider $g\ne 1$. The type of excitation will depend on its conjugacy class. First, consider a group element $g$ where $\pi(g)=1$. We see that only two $\bs B_{p_{\parallel}}$ terms at the ends of the ribbon are violated. Importantly, all the $\bs B_{p_\perp}$ operators touching this ribbon are not violated since the modified group element on each edge is of the form $ \overline{\left(\prod_i k_i\right )} g \left(\prod_i k_i) \right)$, which is the same conjugacy class as $g$. Therefore, they are also projected by $\pi$ to $1$. We therefore see that the resulting excitations are a pair of planons.

Next, consider the case where $\pi(g)\ne 1$. Using the same ribbon operator as above, we find that both $\bs B_{p_\perp}$ and $\bs B_{p_{\parallel}}$ are violated. Therefore, the resulting flux is a loop excitation. This loop excitation is fully mobile, as we can also bend the loop into the $xy$ plane by left multiplying with $\pi(g)$ on $z$ edges and appropriately conjugating $\pi(g)$ as we expand the loop.

Therefore, we have constructed flux excitations (violation of plaquette terms), whose type depends on the conjugacy class. In general, charge excitations can be created as well, forming either a dyon (bound state of charge and flux planons) or a bound state of a point charge and a flux loop. Further analysis is required to find the proper basis for these excitations.

To conclude this section, we reinterpret the local Wilson operators as a product of closed trajectories of the charge and flux operators. The projector $\bs B_{p_\parallel}$ can be decomposed in terms of characters of $G$: $\chi^{\bs \mu,G} = \text{Tr} ( \Gamma^{\bs \mu,G})$ since
\begin{align}
    \delta_{1,g_1g_2\bar g_3\bar g_4} = \frac{1}{|G|} \sum_{\bs\mu} d_{\bs \mu} \chi^{\bs \mu,G} (g_1g_2\bar g_3\bar g_4)
\end{align}
Physically, each term in the sum denotes a closed loop of the gauge charge for each irrep $\bs \mu$. Similarly, $\bs B_{p_\perp}$ can be decomposed into characters of $Q$ since
\begin{align}
    \delta_{1,q_1'q_2'\bar q_3' \bar q_4'} = \frac{1}{|Q|} \sum_{\bs\nu} d_{\bs \nu} \chi^{\bs \nu,Q} (q_1'q_2'\bar q_3' \bar q_4').
\end{align}
Here, we sum over irreps $\bs \nu$ of $Q$, which in turn, can be pulled back to the mobile irreps of $G$. Therefore, each term in the sum denotes a closed loop of all the mobile charges.

\section{3-foliated $(G,N)$ gauge theory}\label{sec:generalGN}
In this section, we give the commuting projector Hamiltonian that realizes the 3-foliated $(G,N)$ gauge theory. We first introduce the necessary terminology required to define the Hamiltonian in Sec. \ref{sec:fracton}. Notably, we review the factor system corresponding to a group extension in Sec. \ref{sec:factor}. In Sec. \ref{sec:derivation}, we derive this Hamiltonian by gauging a paramagnet with a particular mix of global and subsystem symmetries, which we call a 2-subsystem symmetry.

Our discussion here will focus on the case where the subsystem symmetry consists of three planes on a cubic lattice (generalizing the X-cube model\cite{VijayHaahFu2016}). Nevertheless it is possible to write down similar models
with other subsystem symmetries, including those with fractal support, though a thorough discussion of hybrid orders with fractal subsystem symmetries is beyond the scope of this work.

\subsection{The lattice model}\label{sec:fracton}

\begin{figure}[t!]
    \centering
    \includegraphics[scale=0.8]{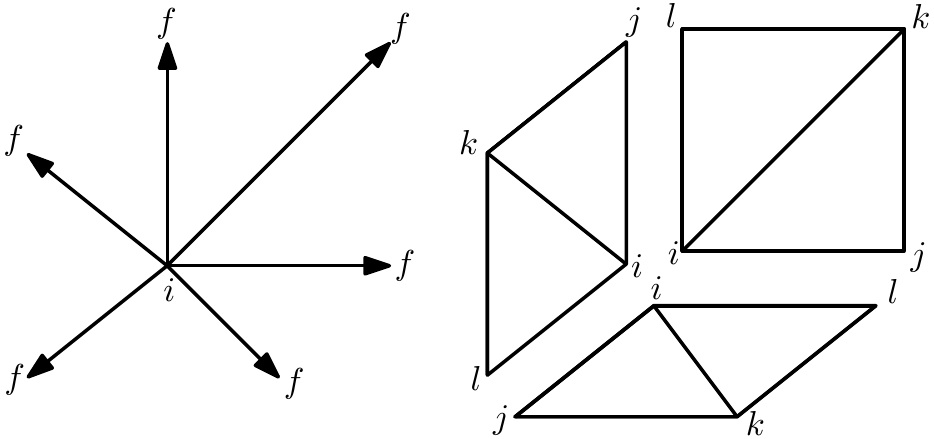}
    \caption{{\bf Description of the Lattice for the 3-foliated $(G,N)$ gauge theory:} Diagonal edges are added to each plaquette in the cubic lattice. \emph{Left}: each edge $e=(if)$ is oriented, pointing outward from an ``initial" vertex $i$ towards a ``final" vertex $f$. \emph{Right}: ordering of vertices for each square plaquette $p=(ijkl)$}
    \label{fig:ordering_general}
\end{figure} 

Our model is defined on a cubic lattice with additional diagonal edges as shown in Figure \ref{fig:ordering_general}. We place an $N$ degree of freedom $\mathbb C[N]$ on each (square) plaquette and a $Q=G/N$ degree of freedom $\mathbb C[Q]$ on each edge of the lattice.
The Hamiltonian takes the following form, 
\begin{align}
H=& -\sum_v \bs A^{G}_v   -\sum_c\sum_{r=x,y,z} \bs B^N_{c,r} - \sum_{\nablapic} \bs B^Q_{{\nablapic}},
\label{equ:GNHam}
\end{align}
where $\bs A^{G}_v$, $\bs B^N_{c,r}$ and $\bs B^Q_{{\nablapic}}$ are commuting projectors defined on each vertex $v$, cube $c$, and triangle ${\nablapic}$ in the lattice, respectively.

The terms in the Hamiltonian are such that
\begin{enumerate}
    \item When $N=\ZZ_1$, $\bs B^N_{c,r} = \mathbbm 1$ and $\bs A^G_v$ and $\bs B^Q_{\nablapic}$ are respectively the vertex and plaquette terms of the 3D quantum double model with gauge group $G=Q$.
    \item When $Q=\ZZ_1$, $\bs B^Q_{\nablapic} = \mathbbm 1$ and $\bs A^G_v$ and $\bs B^N_{c,r}$ are respectively the vertex and cube terms of the X-cube model with gauge group $G=N$.
\end{enumerate}

It is helpful to describe qualitatively the aspects of this $(G,N)$ gauge theory. The projectors $\bs B^Q_{\nablapic}$ and $\bs B^N_{c,r}$ are delta functions which enforces the gauge constraint (we only enforce them energetically in our Hamiltonian, but they can also be strictly enforced if needed). The vertex term can be decomposed as
\begin{align}
    \bs A^{G}_v  =& \frac{1}{|G|}\sum_{g \in G }\bs  A^{g}_v.
\end{align}
By factoring a group element $g\in G$ into a pair $(n,q)$ where $n\in N$ and $q \in Q$, the vertex term $\bs A^{g}_v$ can be thought of as the vertex term in the 3D $Q$ quantum double model corresponding to the group element $q$, but dressed with controlled operators in such a way that the multiplication of $\bs A^{g}_v$ obeys the group law in $G$. That is, for $g,h \in G$,
\begin{align}
\bs A^{g}_v  \bs A^{h}_v = \bs A^{gh}_v
\label{equ:grouplawG}
\end{align}

There are various excitations in the model, either labeled by representations of the group $G$ (charges) or conjugacy classes of $G$ (fluxes) with different mobilities. They are measured by the violations of terms in this commuting projector Hamiltonian. First,
charge excitations are violations of $\bs A^{G}_v=1$. From the group law above, it follows that charge excitations must transform as irreps of $G$. And non-abelian charges are those transform under higher-dimensional irreps of $G$. Furthermore, the model has the property that for group elements in the normal subgroup $N$, the product of vertex terms over individual planes is one
\begin{align}
    \prod_{v \in \text{plane}} \bs A^g_v=1, ~~~ \forall g\in \iota(N).
\end{align}
This conservation law implies that any excitation that transform non-trivially under the normal subgroup $N$ must be a fracton. Otherwise, that excitation is fully mobile.

 Second, the flux term $\bs B^Q_{\nablapic}$ is the same as that of the $Q$ quantum double model. A violation of this flux term represents a mobile flux loop. Finally, $\bs B^N_{c,r}$ is a modified flux term of the X-cube model with gauge group $N$, whose violiations are lineon excitations. 

In order to describe precisely the terms in the Hamiltonian, we will first review the necessary machinery. First, we review factor systems corresponding to group extensions, and define corresponding operators used to define the projectors in our Hamiltonian.

\subsubsection{Factor systems}\label{sec:factor}
In the following, all groups are defined with the usual multiplication operation, $1$ denotes the identity element, and the inverse of $g\in G$ is denoted $\bar g$.

Given a group $G$ and an abelian normal subgroup $N \triangleleft G$, one has an exact sequence
\begin{equation}
 1 \rightarrow N  \xrightarrow[]{\iota} G  \xrightarrow[]{\pi} Q \rightarrow 1
\end{equation}
where $\iota$ is injective and $\pi$ is surjective. Such a group extension is determined by a factor system, which consists of two pieces of data
\begin{enumerate}
    \item A map $\sigma:Q\rightarrow \text{Aut}(N)$. That is for each $q \in Q$, $\sigma^q$ defines an automorphism on $N$.
    \item A cocycle $\omega:Q^2\rightarrow N$ which represents a cohomology class  $[\omega] \in H^2_\sigma(Q,N)$.
\end{enumerate}
 Furthermore, we will also make the following assumptions, which for each factor system can always be chosen
\begin{enumerate}
\item $\sigma^1$ acts as the identity automorphism.
\item A canonical choice of ``normalized'' cocycles where $\omega(1,q) = \omega(q,1)=1$.
\end{enumerate}

Let us now explain the group extension. Given a factor system, any $g \in G$ can be uniquely labeled as a pair $(n,q)\equiv \fs{n}{q}$ for some $n \in N$, and $q\in Q$. The group law is given by
\begin{align}
    \fs{n_1}{q_1} \cdot \fs{n_2}{q_2} = \fs{\omega(q_1,q_2)n_1\aut{q_1}{n_2}}{q_1q_2}
\end{align}
Here, associativity of group multiplication is guaranteed by the fact that $\omega$ satisfies the so-called cocycle condition
\begin{equation}
    \aut{q_1}{\omega(q_2,q_3)} \bar \omega(q_1q_2,q_3)  \omega(q_1,q_2q_3) \bar \omega(q_1,q_2)=1 .
    \label{equ:cocyclecondition}
\end{equation}
Here, $\bar \omega$ is the inverse cocycle of $\omega$.

The maps $\iota$ and $\pi$ in the short exact sequence are given by
\begin{align}
  \iota(n) &= \fs{n}{1}\\
  \pi(\fs{n}{q})&= q
\end{align}
So it is clear that $\pi \circ \iota = 1$, the identity element in $Q$. For convenience, we also define a map $t:G \rightarrow N$ as
\begin{align}
    t(g) =t(\fs{n}{q}) \equiv n
    \label{equ:tdef}
\end{align}
so that we can express a group element $g$ as 
\begin{align}
    g = \fs{n}{q}= \fs{t(g)}{\pi(g)}.
\end{align}
Note that unlike $\pi$, the map $t$ is not a homomorphism.

The maps $\pi$ and $t$ satisfy the following useful identities which can be proven from the definitions. They will be useful for the gauging process.
\begin{align}
 \pi(gg_v) &=q q_v\label{equ:pimaponproduct}\\
  \pi(g_v\bar g) &= q_v\bar q\\
  t(gg_i\bar g_f)&= \bar \omega(q,q_i\bar q_f) n \aut{q}{t(g_i\bar g_f)} , \label{equ:Lactoni}\\
 t(g_i\overline{gg_f}) &= t(g_i\bar g_f)\bar \omega(q_i\overline{qq_f},q) \aut{q_i\overline{qq_f}}{\bar n},  \label{equ:Lactonf}\\
 t(g_i (\fs{n}{1}) \bar g_f)&= \aut{q_i}{n} t(g_i\bar g_f).
 \label{equ:multiply_n_in_t}
\end{align}

\subsubsection{Definition of operators}\label{sec:operatordef}

Next, we proceed to define operators in the Hilbert space of our model.  We consider a cubic lattice with an additional diagonal edge added to each plaquette on the cubic lattice as shown in Fig. \ref{fig:ordering_general}. We place the group algebra $\CC[N]$ on each square plaquette of the cubic lattice, and $\CC[Q]$ on each edge. First of all, we can define the usual left and right multiplication operators as in the quantum double model
\begin{align}
L_p^n \ket{n_p} &= \ket{nn_p}, \\
R_p^n \ket{n_p} &= \ket{n_p\bar n}, \\
L_e^q \ket{q_e} &= \ket{qq_e}, \\
R_e^q \ket{q_e} &= \ket{q_e\bar q} ,
\end{align}
which acts purely either on the $N$ d.o.f. (each plaquette) or the $Q$ d.o.f. (each edge). In addition, the factor system allows us to define additional operators. First, for a fixed $q \in Q$, we can define 
\begin{align}
      \Sigma^q_p \ket{n_p} &=   \ket{\aut{q}{n_p}}
\end{align}
which applies the automorphism $\sigma^q$ to $n$. Next, as a slight abuse of notation, we define
\begin{align}
    \Sigma^e[n] \ket{q_e} &=  \aut{q_e}{n} \ket{q_e}\\
    \Omega(e,q) \ket{q_e} &=  \omega(q_e,q) \ket{q_e}\\
     \Omega(q,e) \ket{q_e} &=  \omega(q,q_e) \ket{q_e}
\end{align}
which for a fixed $n$ or $q$ reads off the $Q$ d.o.f. $q_e$ on edge $e$ and constructs $N$ group elements $\aut{q_e}{n}$, $\omega(q_e,q)$, or $\omega(q,q_e)$. The above are not operators per se, but can be used to define controlled operators with $Q$ d.o.f. as the control site and $N$ d.o.f as the target site by combining them with left and right multiplication. For example, the operator $L_p^{\Sigma_e[n]}$ is defined as
\begin{align}
    L_p^{\Sigma^e[n]} \ket{n_p,q_e} =  L_p^{\sigma^{q_e}[n]} \ket{n_p,q_e} = \ket{\sigma^{q_e}[n]n_p,q_e}
\end{align}
which reads the $Q$ d.o.f. on site $e$, $q_e$, permutes $n$ by $\sigma^{q_e}$ then left multiplies the result to the plaquette $p$. For completion, we define the operators used in the paper according to this notation

\begin{figure}[t!]
    \centering
    \includegraphics[scale=1]{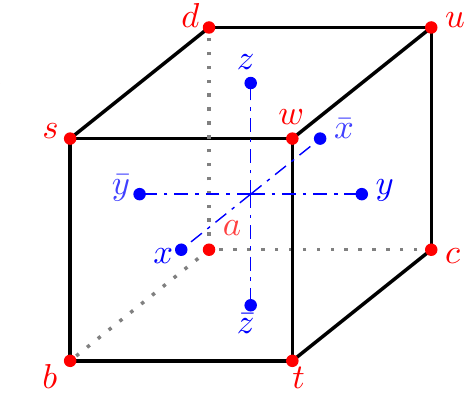}
    \caption{Label for the vertices (red) and faces (blue) for each cube used to define the projector $\bs B^N_{c,r}$.}
    \label{fig:Xcubeconstraint}
\end{figure}

\begin{align}
R_p^{\Sigma^e[n]} \ket{n_p,q_e} &=   R_p^{\sigma^{q_e}[n]} \ket{n_p,q_e} = \ket{n_p\sigma^{q_e}[\bar n],q_e}\\
L_p^{\Omega(e,q)} \ket{n_p,q_e} &= L_p^{\omega(q_e,q)} \ket{n_p,q_e} = \ket{\omega(q_e,q)n_p,q_e}\\
R_p^{\Omega(e,q)} \ket{n_p,q_e} &= R_p^{\omega(q_e,q)} \ket{n_p,q_e}=  \ket{n_p\bar\omega(q_e,q),q_e}\\
L_p^{\Omega(q,e)} \ket{n_p,q_e} &= L_p^{\omega(q,q_e)} \ket{n_p,q_e} = \ket{\omega(q,q_e)n_p,q_e}\\
R_p^{\Omega(q,e)} \ket{n_p,q_e} &= R_p^{\omega(q,q_e)} \ket{n_p,q_e}=  \ket{n_p\bar\omega(q,q_e),q_e}
\end{align}
Lastly, we define the following controlled operator
\begin{align}
    \Sigma^e_p \ket{n_p,q_e} &=   \ket{\aut{q_e}{n_p},q_e}
\end{align}
which reads off $q_e$ and applies the corresponding permutation $\sigma^q_e$ to $n_p$.

\subsubsection{Definition of projectors}
Having defined the operators, we can now define our model. First, we define a local ordering of the vertices to each edge $e = (if )$ and each square plaquette $p = (ijkl)$ as shown in Fig. \ref{fig:ordering_general}, and denote $e \rightarrow v$ ($e \leftarrow v$) as incoming (outgoing) edges towards (from) the vertex $v$. The projectors are given by
\begin{widetext}
\begin{align}
\bs A^{G}_v  &= \sum_{g \in G }\bs  A^{g}_v\\
\bs  A^{g}_v  &=\prod_{p|v=j,l}  R_p^{\Omega ((iv) ,q) \Sigma^{(iv)}[n]} \times \prod_{p|v=k}  L_p^{\Omega ((iv) ,q) \Sigma^{(iv)}[n]} \times \prod_{e \leftarrow v} R^{q}_e \prod_{e \rightarrow v}L^q_e \times \prod_{p|v=i} L_p^{\bar \Omega(q,(vj))  \Omega(q,(vk)) \bar \Omega(q,(vl))n} \Sigma^q_p,\\
\bs B^Q_{\nablapic} &= \sum_{\{q\}} \delta_{1,q_{ad}q_{du}\bar q_{au}} \ket{\{q\}}\bra{\{q\}}, \ \ \ \ \ \ \ \ ;{\nablapic} = (adu) \\
 \bs  B^N_{c,x}&=\sum_{\{n,q\}}  \delta_{1,\omega(q_{ad},q_{ds})\bar \omega(q_{ad},q_{dw})\omega(q_{ad},q_{du})\bar \omega(q_{ac},q_{ct})\omega(q_{ac},q_{cw})\bar \omega(q_{ac},q_{cu}) \aut{q_{ac}}{n_y} \aut{q_{ad}}{\bar n_{ z}} \bar n_{\bar y} n_{\bar z}} \ket{\{n,q\}} \bra{\{n,q\}}\\
\bs  B^N_{c,y}&=\sum_{\{n,q\}}  \delta_{1,\omega(q_{ab},q_{bt})\bar \omega(q_{ab},q_{bw})\omega(q_{ab},q_{bs})\bar \omega(q_{ad},q_{du})\omega(q_{ad},q_{dw})\bar \omega(q_{ad},q_{ds}) \aut{q_{ad}}{n_z} \aut{q_{ab}}{\bar n_{x}} \bar n_{\bar z} n_{\bar x}} \ket{\{n,q\}} \bra{\{n,q\}}\\
\bs  B^N_{c,z}&=\sum_{\{n,q\}}  \delta_{1,\omega(q_{ac},q_{cu})\bar \omega(q_{ac},q_{cw})\omega(q_{ac},q_{ct})\bar \omega(q_{ab},q_{bs})\omega(q_{ab},q_{bw})\bar \omega(q_{ab},q_{bu}) \aut{q_{ab}}{n_x} \aut{q_{ac}}{\bar n_{y}} \bar n_{\bar x} n_{\bar y}} \ket{\{n,q\}} \bra{\{n,q\}}
    \end{align}
    \end{widetext}
Here, the edges used to define the $\bs B^N_{c,r}$ operators are those connecting two vertices as labeled in Figure \ref{fig:Xcubeconstraint}. We further note that the three $\bs B^N_{c,r}$ operators are just $C_3$ rotations of each other around the $(1,1,1)$ axis, which is achieved via cyclically permuting the vertices $b,c,d$, vertices $s,t,u$, plaquettes $x,y,z$, and plaquettes $\bar x,\bar y,\bar z$

With the explicit expressions, let us verify the claims made earlier 
\begin{enumerate}
    \item When $N=\ZZ_1$, the only elements of $N$ are $1$ and so $\bs B^N_{c,r} = \mathbbm 1$. The vertex term reduces to
    \begin{align}
        \bs A^g_v = \frac{1}{|G|}\prod_{e \leftarrow v} R^{q}_e \prod_{e \rightarrow v}L^q_e
    \end{align} which is just the vertex term of the 3D quantum double model with gauge group $G=Q$.
    \item When $Q=\ZZ_1$, $\bs B^Q_{\nablapic} = \mathbbm 1$. The vertex and cube terms reduce to
    \begin{align}
        \bs A^g_v &=\prod_{p|v=j,l}  R_p^{n} \times \prod_{p|v=i,k} L_p^{n} \\
        \bs  B^N_{c,x}&=\sum_{\{n\}}  \delta_{1,n_y \bar n_{ z} \bar n_{\bar y} n_{\bar z}} \ket{\{n\}} \bra{\{n\}}\\
        \bs  B^N_{c,y}&=\sum_{\{n\}}  \delta_{1,n_z \bar n_{ x} \bar n_{\bar z} n_{\bar x}} \ket{\{n\}} \bra{\{n\}}\\
        \bs  B^N_{c,z}&=\sum_{\{n\}}  \delta_{1,n_x \bar n_{y} \bar n_{\bar x} n_{\bar y}} \ket{\{n\}} \bra{\{n\}}
    \end{align} 
   which are those of the X-cube model with gauge group $G=N$.
\end{enumerate}

Let us now analyze the terms of the Hamiltonian in more detail. First, consider the vertex term $\bs A^g_v$.  When restricted to the subgroup $N$ (i.e. those of the form $g= \fs{n}{1}$), the vertex term $\bs A^g_v$ is that of the X-cube model with gauge group $N$. Furthermore, when restricted to group elements of the form $g= \fs{1}{q}$, the vertex term $\bs A^g_v$ can be interpreted as the vertex term of the $Q$ quantum double model dressed with certain combinations of controlled operators. These operators guarantee the correct group multiplication of $\bs A^g_v$ as given in Eq. \eqref{equ:grouplawG} .

Moving on to the flux terms, the plaquette term $\bs B^Q_{\nablapic}$ is the same as that in the $Q$ quantum double model. A violation of this plaquette represents a mobile flux loop labeled by a conjugacy class of $G$ whose image under $\pi$ is a non-trivial conjugacy class of $Q$. The cube term $\bs B^N_{c,r}$ is a modified flux term of the X-cube model with gauge group $N$. Violations of such terms are lineons labeled by the remaining conjugacy classes of $G$.

\subsection{Derivation of the 3-foliated $(G,N)$ gauge theory}\label{sec:derivation}

In this section, we provide a derivation of the 3-foliated $(G,N)$ gauge theory. We will start by introducing a paramagnet with a 2-subsystem symmetry, which we will rigorously define. Then, we will proceed to gauge the 2-subsystem symmetry, which can be done in two steps. First, we gauge the $N$ subsystem symmetry, which is a normal subgroup of $(G,N)$ to obtain an X-cube model with gauge group $N$ enriched by a global symmetry $Q=G/N$. Then, we will gauge $Q$ to obtain the $(G,N)$ gauge theory. The equivalent gauging procedure for the quantum double model is reviewed in detail in Appendix \ref{sec:QDmodel}.

\subsubsection{2-Subsystem Symmetry} 

We will begin with a proper definiton of a 2-subsytem symmetry. A brief discussion of abelian 2-subsystem symmetries has been presented in Ref. \onlinecite{TJV1}.

A 2-subsystem symmetry can be defined given the following data
\begin{enumerate}
    \item A global symmetry group $G$ 
    \item A normal subgroup $N \triangleleft G$
    \item The type of region on which $N$ acts as a subsystem symmetry (e.g. 1-foliated, 3-foliated, fractal,...)
\end{enumerate}

Note that first two data is equivalent to specifying $N$, $Q$ and the factor system $\sigma$ and $\omega$.

If we ignore the geometrical input of the type of subsystem symmetry, then $(G,N)$ is a group given uniquely by the group extension
\begin{equation}
1 \rightarrow N^L  \xrightarrow[]{} (G,N)  \xrightarrow[]{} Q \rightarrow 1
\label{equ:exact2}
\end{equation}
where $N^L \equiv N_1 \times \cdots \times N_L$ is a product of $L$ identical copies of $N$, which represents the number of independent subsystem symmetries.
The group extension above can be specified by the map $\sigma': Q\rightarrow \text{Aut}(N^L)$ which acts as $\sigma$ identically on all copies of $N$ in $N^L$, and a representative cocycle $\omega'$ of $H^2(Q,N^L)$ chosen to be $ \omega' = \prod_{l=1}^L \iota_l \circ \omega $ where $\iota_l: N \rightarrow N^L$ is the embedding of $N$ to the $l^\text{th}$ copy of $N^L$.

The 2-subsystem symmetry can be realized as the following on a lattice model. First, we define the local Hilbert space on each vertex of the lattice to be the group algebra $\mathbb C[G]$, whose basis vectors are denoted $\ket{g}$ for $g \in G$. The onsite symmetry action of $G$ acts via right multiplication
\begin{align}
    R^g_v: \ket{h_v} \rightarrow \ket{h_v \bar g}
\end{align}
The 2-subsystem symmetry can now be defined as
\begin{align}
\begin{split}
       \text{Global symmetry $G$:} & \ \ \ \prod_v R^g_v\\
    \text{Subsystem symmetry $N$:} & \ \ \ \prod_{v \in \text{sub}} R^n_v 
\end{split}
\label{equ:2subsystemaction}
\end{align}
Here, ``sub" is a subregion specified by the geometric input of the subsystem symmetry.

In passing, we would like to make a comparison to 2-group symmetries\cite{KapustinThorngren2017,DelcampTiwari18,CordovaDumitrescuIntriligator19,BeniniCordovaHsin19}, which can be thought of as a particular extension of global (0-form) symmetries by 1-form symmetries (and whose name we were inspired from). In particular, one could ask why we expect the input data for 2-subsystem symmetry to depend on a 2-cocycle rather than a 3-cocycle as in 2-groups. Here, we will give a physically motivated answer from the perspective of fractionalization. Recall that in 2D, gauging a normal subgroup $N$ of a paramagnet with $G$ symmetry, results in an $N$ topological order where the symmetry fractionalization on the charge anyons are determined by a cohomology class $H^2_\sigma(Q,N)$, which determines the group extension\cite{BarkeshliBondersonChengWang2019,Hermele2014,TarantinoLinderFidkowski2016}. Similarly, for 2-groups, after gauging the 1-form symmetry $N$ the resulting abelian topological order is an SET where the remaining 0-form global symmetry $Q$ fractionalizes on the loop excitations according to the cohomology class  $H^3_\sigma(Q,N)$.\footnote{for an explicit lattice calculation, see for example Ref. \onlinecite{ChenEllisonTantivasadakarn20}.} A way to interpret the different fractionalization classes is to view the corresponding string/membrane operators of the anyons or loop excitations as attached with different 1d or 2d ``$N$-valued SPTs protected by $Q$". In this way, the fractionalization can be interpreted as an anomaly coming from the boundary of the attached SPT phase \cite{ElseNayak2014}. 

Returning to the case for 2-subsystem symmetry, after gauging the subsystem symmetry $N$, the gauge charges are fractons, which are 0-dimensional. Therefore, the fractionalization class should be thought of as the end points of a 1D SPT, and therefore it is reasonable to expect that this is captured via a 2-cocycle in $H^2_\sigma(Q,N)$.

\subsubsection{2-subsystem paramagnet}\label{sec:Isingmodel3D}

To begin the derivation, we will start with a cubic lattice where each vertex is $\CC[G]$. In addition to the paramagnetic term, we will write down additional kinetic terms that respects the 2-subsystem symmetry.

On a cubic lattice with extra diagonal edges as shown in Fig. \ref{fig:ordering_general}, we place the group algebra $\mathbb C[G]$ on each vertex. The Hamiltonian we will consider is

    \begin{align}
H=-\sum_v L_v-h_\text{E}\sum_e \Delta_{e} -h_\text{P}\sum_p \Delta_p ,
\end{align}
which consists of the paramagnetic term and two types of kinetic terms $\Delta$. For each edge $e= (if)$ and plaquette $ p= (ijkl)$, we define
\begin{align}
\Delta_{e}&=\sum_{q_i,q_f}\delta_{1,q_i\bar q_{f}}\ket{q_i,q_{f}}\bra{ q_i,q_{f}}.          \label{equ:Deltae3D}\\
\Delta_{p}&=\sum_{g_i,g_j,g_k,g_l}\delta_{1,t(g_i\bar g_j) \overline{t(g_i\bar g_k)} t(g_i\bar g_l) } \nonumber\\
      &\ket{g_i,g_j,g_k,g_l}\bra{ g_i,g_j,g_k,g_l}
          \label{equ:Deltap3D}
\end{align}
where $t$ is defined in Eq. \eqref{equ:tdef}. It is apparent that $\Delta_e$ commutes with the 2-subsystem symmetry Eq. \eqref{equ:2subsystemaction}, and we can see that $\Delta_p$ commutes with the global $G$ symmetry, since right multiplication on all sites leaves terms of the form $t(g_i\bar g_j)$ invariant. Therefore, we only need to check that $\Delta_p$ commutes with the $N$ planar symmetry. As an example, let us consider the case where the planar symmetry acts only on sites $i$ and $j$ of the plaquette $p$. $t(g_i\bar g_j)$ is invariant, and using Eq. \eqref{equ:multiply_n_in_t}, we see that the contributions coming from $\overline{t(g_i\bar g_k)}$ and $t(g_i\bar g_l)$ are respectively $\aut{q_i}{\bar n}$ and $\aut{q_i}{n}$, which cancel. Similarly, one can confirm that $\Delta_p$ is also invariant when the symmetry acts only on sites $j$ and $k$, $k$ and $l$, or $l$ and $i$. Therefore, our Hamiltonian commutes with the 2-subsystem symmetry.

There are some remarks to be made about our Hamiltonian. First, the Hamiltonian does not have a planar symmetry for the full group $G$, which avoids the problem of having an additional local symmetry at each site given by the commutator subgroup $[G,G]$ when $G$ is non-abelian. To see this, we note that the edge kinetic term $\Delta_e$ does not commute with a planar symmetry $g$ unless it is of the form $g=\fs{n}{1}$. This is because for a given plane, one can consider an edge normal to the plane such that the symmetry acts on only one site of $\Delta_e$ (say site $i_e$). In such case, the delta function will transform as
\begin{align}
    \delta_{1,q_i\bar q_f} \rightarrow \delta_{1,q_i\bar q\bar q_f}
\end{align}
which is not invariant. Second, the constraint in the delta function for $\Delta_p$ is not equal to $t(g_i\bar g_j g_k \bar g_l)$. In fact, one can check that such a term does not commute with the 2-subsystem symmetry. Lastly, the Hamiltonian has additional symmetries coming from left multiplication, but they will not be important for our discussion, as we will gauge only the right multiplication symmetries.

\subsubsection{Gauging $N$ planar symmetries}\label{sec:SEF}

We will now gauge the $N$ planar symmetries (acting via right multiplication), and argue that we obtain a Symmetry-Enriched Fracton (SEF) order, where the fracton model is in the same phase as the X-cube model with gauge group $N$, but is enriched by a global symmetry $Q$. We demonstrate the permutation and fractionalization of the excitations of the SEF in Appendix \ref{app:GaugingNFracton}. 

To gauge the $N$ planar symmetries, we define the following ``gauging map", which projects each vertex to a $Q$ d.o.f. via $\pi$ and maps the expression in the delta function of $\Delta_p$ to an $N$-plaquette d.o.f.. That is, for the four vertices surrounding a plaquette $p$, we map
\begin{equation}
   \raisebox{-.5\height}{\begin{tikzpicture}
\node[label=center:$g_i$] (1) at (0,0) {};
\node[label=center:$g_l$] (2) at (0,\l) {};
\node[label=center:$g_k$] (3) at (\l,\l) {};
\node[label=center:$g_j$] (4) at (\l,0) {};
\draw (1) -- (2){};
\draw (2) -- (3){};
\draw (1) -- (4) {};
\draw (4) -- (3){};
\end{tikzpicture}} \rightarrow
   \raisebox{-.5\height}{\begin{tikzpicture}
\node[label=center:$q_i$] (1) at (0,0) {};
\node[label=center:$q_l$] (2) at (0,\l) {};
\node[label=center:$q_k$] (3) at (\l,\l) {};
\node[label=center:$q_j$] (4) at (\l,0) {};
\draw[] (1) -- (2){};
\draw[] (2) -- (3){};
\draw[] (1) -- (4) {};
\draw[] (4) -- (3){};
\node[label=center:$n_p$]  at (\l/2,\l/2) {};
\end{tikzpicture}} 
\label{equ:gaugingmap3D}
\end{equation}
where 
\begin{align}
    n_p=&t(g_i\bar g_j) \overline{t(g_i\bar g_k)} t(g_i\bar g_l) 
\end{align}
Note that the expression of $n_p$ is just the constraint in the kinetic term Eq. \eqref{equ:Deltap3D}. The reason this map can be thought of as gauging the planar symmetries because $n_p$ is invariant under the planar symmetry action. Therefore, after the mapping only the global symmetry $Q=G/N$ is left as a physical symmetry acting on the system.

Let us work out how the operators are mapped. The Ising term $\Delta_e$ is unchanged, while $\Delta_p$ gets mapped to $T^1_p \equiv\ket{1}\bra{1}_p$. Next, we must compute how $L^g_v$ maps. For each plaquette $p$ whose corner contains $v$, we identify whether $v$ constitutes as the vertex $i,j,k$, or $l$ of that plaquette. If $v=i$, then using Eq. \eqref{equ:Lactoni}, we see that under the action of $L^g_v$,
\begin{align}
    L^g_{v}\ket{n_p}= \Ket{\bar \omega(q,q_i\bar q_j) \omega(q,q_i\bar q_k) \bar \omega(q,q_i\bar q_l) n \aut{q}{n_p}}
\end{align}
Using Eq. \eqref{equ:Lactonf}, for $v=k$ we find
\begin{align}
    L^g_{v}\ket{n_p}= \Ket{ \omega(q_i\overline{qq_k},q)  \aut{q_i \bar {qq_k}}{n} n_p }
\end{align}
while for $v=j,l$ we find
\begin{align}
    L^g_{v}\ket{n_p}= \Ket{    n_p \bar \omega(q_i\overline{qq_k},q) \aut{q_i \bar {qq_k}}{\bar n} }
\end{align}
Therefore, we conclude that
\begin{align}
    L^{g}_v \rightarrow A^g_v =&\prod_{p|v=j,l}  R_p^{\Omega (i\bar v ,q) \Sigma^{i\bar v}[n]} \times \prod_{p|v=k}  L_p^{\Omega (i\bar v ,q) \Sigma^{i\bar v}[n]}\nonumber\\
    &\times L^q_v \times \prod_{p|v=i} L_p^{\bar \Omega(q,v\bar j)  \Omega(q,v\bar k) \bar \Omega(q,v\bar l)n} \Sigma^q_p.\label{equ:AgvinSEF}
\end{align}
where in the above we have defined
\begin{align}
\Sigma^{i\bar v}[n] \ket{q_i,q_v} &= \sigma_p^{q_i\bar q_v }[n]\ket{q_i,q_v},\\
\Omega(i\bar v,q) \ket{q_i,q_v} &= \omega(q_i\bar q_v,q)\ket{q_i,q_v},\\
\Omega(q,v\bar i) \ket{q_i,q_v} &= \omega(q,q_v\bar q_i)\ket{q_i,q_v}.
\end{align}

Finally, we impose the constraint which arises from gauging. Consider a cube with plaquettes and vertices depicted in Figure \ref{fig:Xcubeconstraint}. We notice that for each plaquette $p$, if we define
\begin{align}
 \tilde n_p&= \aut{\bar q_i} { \omega(q_i\bar q_j,q_j)\bar \omega(q_i\bar q_k,q_k) \omega(q_i\bar q_l,q_l)n_p }\nonumber\\
 &=\aut{\bar q_i} {n_i}\aut{\bar q_j}{\bar n_j}\aut{ \bar q_k}{n_k}\aut{\bar q_l}{\bar n_l},
 \label{equ:tilden_p}
\end{align}
then we see that the following product of $\tilde n_p$ over four plaquettes around a cube is the identity.
\begin{align}
    \tilde n_{y}\overline{\tilde n_{z}} \ \overline{\tilde n_{\bar y}} n_{\bar z}  =1
\end{align}
Therefore, we can impose the following constraint energetically via
\begin{align}
B^N_{c,x} &=\sum_{\{n,q\}} \delta_{1,  \tilde n_{y}\overline{\tilde n_{z}} \ \overline{\tilde n_{\bar y}} n_{\bar z}} \ket{\{n,q\}}\bra{\{n,q\}}\\
B^N_{c,y} &=\sum_{\{n,q\}} \delta_{1,  \tilde n_{z}\overline{\tilde n_{x}} \ \overline{\tilde n_{\bar z}} n_{\bar x}} \ket{\{n,q\}}\bra{\{n,q\}}\\
B^N_{c,z} &=\sum_{\{n,q\}} \delta_{1,  \tilde n_{x}\overline{\tilde n_{y}} \ \overline{\tilde n_{\bar x}} n_{\bar y}} \ket{\{n,q\}}\bra{\{n,q\}}
    \label{equ:BN3D}
\end{align}

To conclude, the symmetry-enriched fracton Hamiltonian takes the form
\begin{align}
H_\text{SEF} &= -\frac{1}{|G|} \sum_v \sum_{g \in G} A^{g}_v  - \sum_c\sum_{r=x,y,z} B^N_{c,r}\nonumber\\
&- h_\text{E}\sum_e \Delta_e - h_\text{P}\sum_p T^1_p, 
\label{equ:SEF}
\end{align}
where the vertex term $A^{g}_v$ is given in Eq. (\ref{equ:AgvinSEF}), and $B^N_{c,r}$ is given in Eq. (\ref{equ:BN3D}). The Hamiltonian is exactly solvable in the limit $h_\text{E} = h_\text{P}=0$.

In Appendix \ref{app:GaugingNFracton}, we perform a basis transformation that turns the $N$-d.o.f. on each plaquette from $n_p$ to $\tilde n_p$ defined in Eq. \eqref{equ:tilden_p}. In this basis, the $N$-d.o.f. forms the X-cube model with gauge group $N$, and we explicitly demonstrate the symmetry enrichment via the permutation/fractionalization of the excitations.

\subsubsection{Gauging the global $Q$ symmetry}\label{sec:gaugeQ3D}
\begin{table*}[t]
 \caption{Mapping of operators in the $(G,N)$ paramagnet, the SEF ($N$-X-cube order enriched by $Q$), and the $(G,N)$ gauge theory. The definitions of these operators are summarized in Sec. \ref{sec:operatordef}.}
    \centering
    \begin{tabular}{|c|c|c|}
    \hline
   $(G,N)$-Ising & SEF & $(G,N)$ gauge theory \\
     \hline
 $ L^{g}_v $    &$\displaystyle  A^g_v =\prod_{p|v=j,l}  R_p^{\Omega (i\bar v ,q) \Sigma^{i\bar v}[n]} \times \prod_{p|v=k}  L_p^{\Omega (i\bar v ,q) \Sigma^{i\bar v}[n]}$ &$\displaystyle \bs A^g_v=\prod_{p|v=j,l}  R_p^{\Omega ((iv) ,q) \Sigma^{(iv)}[n]} \times \prod_{p|v=k}  L_p^{\Omega ((iv) ,q) \Sigma^{(iv)}[n]} \times \prod_{e \leftarrow v} R^{q}_e $\\
 &$\displaystyle\times L^q_v \times \prod_{p|v=i} L_p^{\bar \Omega(q,v\bar j)  \Omega(q,v\bar k) \bar \Omega(q,v\bar l)n} \Sigma^q_p$& $\displaystyle \times  \prod_{e \rightarrow v}L^q_e \times \prod_{p|v=i} L_p^{\bar \Omega(q,(vj))  \Omega(q,(vk)) \bar \Omega(q,(vl))n} \Sigma^q_p$\\
 \hline
 $\Delta_e$ & $ \Delta_e$ & $T^1_e$ \\
  \hline
 $\Delta_p$ & $ T^1_p$ & $T^1_p$ \\
  \hline
 $1$ & $\displaystyle  B^N_{c,x} =\sum_{\{n,q\}} \delta_{1,  \tilde n_{y}\overline{\tilde n_{z}} \ \overline{\tilde n_{\bar y}} n_{\bar z}} \ket{\{n,q\}}\bra{\{n,q\}}$ & $ \displaystyle \bs B^N_{c,x} =\sum_{\{n,q\}}  \delta_{1,\omega(q_{ad},q_{ds})\bar \omega(q_{ad},q_{dw})\omega(q_{ad},q_{du})\bar \omega(q_{ac},q_{ct})}$ \\
 & & $_{\omega(q_{ac},q_{cw})\bar \omega(q_{ac},q_{cu}) \aut{q_{ac}}{n_y} \aut{q_{ad}}{\bar n_{\bar z}} \bar n_{\bar y} n_{\bar z}} \ket{\{n,q\}} \bra{\{n,q\}}  $ \\
  \hline
 $1$ & $1$ & $\displaystyle \bs B^Q_{\nablapic} = \sum_{\{q\}} \delta_{1,q_{ad}q_{du}\bar q_{au}} \ket{\{q\}}\bra{\{q\}}$\\
  \hline
   $\displaystyle \prod_{v\in \text{plane}} R^n_v$ & $1$ & $ 1$ \\
    \hline
  $\displaystyle \prod_v R^g_v$ & $\displaystyle \prod_v R^{q}_v$ & $ 1$ \\
    \hline
    \end{tabular}
    \label{tab:duality3D}
\end{table*}
We now gauge the global $Q$ symmetry acting by right multiplication via the gauging map $q_i \bar q_f \rightarrow q_e=q_{(if)}$. This maps $\Delta_e \rightarrow T^1_e = \ket{1}\bra{1}_e$. For the vertex term, we obtain
\begin{widetext}
\begin{align}
    A^g_v &\rightarrow \bs  A^{g}_v  =\prod_{p|v=j,l}  R_p^{\Omega ((iv) ,q) \Sigma^{(iv)}[n]} \times \prod_{p|v=k}  L_p^{\Omega ((iv) ,q) \Sigma^{(iv)}[n]} \times \prod_{e \leftarrow v} R^{q}_e \prod_{e \rightarrow v}L^q_e \times \prod_{p|v=i} L_p^{\bar \Omega(q,(vj))  \Omega(q,(vk)) \bar \Omega(q,(vl))n} \Sigma^q_p.
    \end{align}
To gauge $B^N_{c,x}$. We need to first write it explicitly in terms of $n$ and $q$. Starting from 
Eq. \eqref{equ:BN3D},we can substitute the explicit expression of $\tilde n_p$ in Eq. \eqref{equ:tilden_p} and use the cocycle conditions to simplify. The resulting expression is 
\begin{align}
    B^N_{c,x} =\sum_{\{n,q\}} \delta_{1,     \omega(q_a\bar q_d,q_d\bar q_s)\bar \omega(q_a\bar q_d,q_d\bar q_w)\omega(q_a\bar q_d,q_d\bar q_u)\bar \omega(q_a\bar q_c,q_c\bar q_t)\omega(q_a\bar q_c,q_c\bar q_w)\bar \omega(q_a\bar q_c,q_c\bar q_u) \aut{q_a\bar q_c}{n_y} \aut{q_a\bar q_d}{\bar  n_{\bar z}} \bar n_{\bar y} n_{\bar z}} \ket{\{n,q\}}\bra{\{n,q\}}
\end{align}
Therefore, we can now gauge this to
\begin{align}
    B^N_{c,x} &\rightarrow \bs  B^N_{c,x}=\sum_{\{n,q\}}  \delta_{1,\omega(q_{ad},q_{ds})\bar \omega(q_{ad},q_{dw})\omega(q_{ad},q_{du})\bar \omega(q_{ac},q_{ct})\omega(q_{ac},q_{cw})\bar \omega(q_{ac},q_{cu}) \aut{q_{ac}}{n_y} \aut{q_{ad}}{\bar n_{\bar z}} \bar n_{\bar y} n_{\bar z}} \ket{\{n,q\}} \bra{\{n,q\}} 
\end{align}
\end{widetext}
The expression for $\bs  B^N_{c,y}$ and $\bs  B^N_{c,z}$ are similarly obtained or by performing a $C_3$ rotation around the $(1,1,1)$ axis, which simultaneously permutes vertices $b,c,d$ and $s,t,u$, as well as faces $x,y,z$ and faces $\bar x,\bar y,\bar z$.

In addition, we also impose a fluxless condition for every closed triangle ${\nablapic}$. For example, for the triangle ${\nablapic}=(adu)$ in Figure \ref{fig:Xcubeconstraint}, we have
\begin{align}
   \bs B^Q_{\nablapic} = \sum_{\{q\}} \delta_{1,q_{ad}q_{du}\bar q_{au}} \ket{\{q\}}\bra{\{q\}}
\end{align}

To conclude the desired $(G,N)$ gauge theory can be obtained by setting the transverse fields $h_\text{E}=h_\text{P}=0$, which is the Hamiltonian \eqref{equ:GNHam}.

The duality of the operators are summarized in Table \ref{tab:duality3D}. In Appendix \ref{app:3foliatedexamples} we present various examples.

\section{Discussion}
We have systematically studied hybrid fracton orders that emerge from $(G,N)$ gauge theories, in which the $G$ global symmetry and $N$ subsystem symmetries of a trivial, short-range-entangled quantum system are gauged. We conclude by discussing some open questions (see also the discussion in Ref. \onlinecite{TJV1}). 

{\bf \emph{``Topological data" in Hybrid Fracton Phases}:} It remains an open question to completely characterize the topological data of the excitations in the non-Abelian hybrid orders presented here, similar to the data in 3D quantum double models \cite{JiangMesarosRan2014,MoradiWen15,WangWen2015,Delcamp17,LanKongWen18}.  This data includes the quantum dimensions of the non-Abelian excitations, and their fusion and braiding. It is also interesting to study the three-loop braiding statistics  \cite{WangLevin2014,JianQi14,JiangMesarosRan2014,WangLevin2015,WangWen2015,PutrovWangYau2017,ChengTantivasadakarnWang2018,WangChengWangGu19,ChanYeRyu18,ZhouWangWangGu19,ZhangYe20} for hybrid orders and other exotic processes involving a mix of mobile and restricted excitations.  One could perhaps gain insight from the elementary excitations of 3D quantum double models, including their fusion and braiding properties.

{\bf \emph{Non-Abelian Fractons Beyond $(G,N)$ Gauge Theory}:}. We pointed out that as an intermediate step of gauging the $N$ subsystem symmetry, we obtained a fracton model enriched by a global symmetry $Q$. An obvious generalization is to stack a $Q$-SPT before gauging the global symmetry, which would correspond to considering different possible defectification classes $H^4(Q,U(1))$ of the $Q$ defect loops. This stacking could give rise to interesting hybrid models beyond the current construction. It also remains to be shown whether it is possible to obtain non-abelian fractons that transform under a group $G$ that cannot arise from a group extension, such as the group $A_5$. Another potential direction is whether the construction be generalized to beyond groups. For example, whether there are 3-foliated hybrid fracton orders where the description of the fusion and braiding data calls for more general categories as input. 

{\bf \emph{Continuous groups}:} It would be interesting to construct the analog of $(G,N)$ gauge theories where $G$ is continuous, and study the properties of the resulting (presumably gapless) order. An initiative has begun in this direction using a field theoretical approach in Refs. \onlinecite{WangXu19,WangXuYau19,WangYau19}, where we believe the corresponding 2-subsystem symmetry is $(O(2),U(1))$.

\textbf{Note Added:} While this paper was in preparation, we recently learnt of a work by Tu \& Chang\cite{TuChang21}), which also constructs similar models with non-abelian fractons by gauging a semidirect product of global and subsystem symmetries. The models they consider correspond to the specific case of a split group extension in our work, and can be labeled as $(S_3,\ZZ_3)$ and $(D_4,\ZZ_2^2)$ 3-foliated gauge theories in our terminology.

\begin{acknowledgments}
The authors are grateful to Yu-An Chen and Tyler Ellison for very helpful discussions on symmetry fractionalization and to Ashvin Vishwanath and Yahui Zhang for interesting discussions on realizing fracton models from condensations. The authors thank David Aasen, Kevin Slagle, Ho Tat Lam, Shu-Heng Shao, and Cenke Xu for stimulating discussions. NT is supported by NSERC. WJ is supported by the Simons Foundation. This work was also partially supported by the Simons Collaboration on Ultra-Quantum Matter, which is a grant from the Simons Foundation.  
\end{acknowledgments}

\appendix

\section{Review of Gauging and the Quantum Double model}\label{sec:QDmodel}
In this Appendix, we review the quantum double model of a group $G$ denoted $\mathcal D(G)$ and the process of gauging. We first derive the model by gauging a paramagnet with $G$ symmetry in \ref{sec:GaugeG}. In  \ref{app:factored} - \ref{sec:gaugeQ2D}, we describe an alternative two-step gauging process. First, we ``decompose'' $G$ into a normal subgroup $N$ and its quotient group $Q$ using the factor system of the corresponding group extension in Section \ref{sec:factor}. In \ref{sec:GaugeN2D} we gauge the $N$ global symmetry to obtain an $N$-topological order enriched by the quotient group $Q$. Finally, in \ref{sec:gaugeQ2D} we further gauge $Q$ to recover a similar model which has an identical ground state to $\mathcal D (G)$. As an aside, \ref{app:GaugingN} demonstrates symmetry permutation and fractionalization of the SET obtained as an intermediate step from \ref{sec:GaugeN2D} for some basic groups.

To setup some terminology, we will work with a directed graph, and define $e \rightarrow v$ and $e \leftarrow v$ to denote the set of all edges that enter and exit a vertex $v$, respectively. For each edge $e$, we define $i_e$ and $f_e$ to be the initial and final vertices of each edge, respectively. When it is clear from context, the subscript $e$ might be dropped. $O_e$ denotes the orientation of each edge with respect to a clockwise loop going around each plaquette. Finally, we define a $G$ degree of freedom (d.o.f) to mean a local Hilbert space $\CC[G]$, which is situated of either the vertex or edge of the graph. The basis of each $G$ d.o.f. is denoted by group elements $\ket{g}$ for each $g \in G$, where we will represent the identity element of $G$ as $1$ to not overlap with the symbol for the edge $e$. We will also note the inverse of $g$ by $\bar g$.

\subsection{$\mathcal D(G)$ from gauging $G$}\label{sec:GaugeG}
The quantum double model of a group $G$ is defined on an arbitrary directed graph. For simplicity, we will specialize to a square lattice in 2D with $G$ d.o.f. on each edge $e$. 

First define left and right multiplication operators
\begin{align}
L^g_e \ket{h_e} &= \ket{gh_e} & R^g\ket{h_e} = \ket{h_e\bar g}
\end{align}
and the projector to a group element $g$
\begin{align}
    T^g_e \ket{h_e} = \delta_{g,h_e} \ket{h_e}
    \label{eqn:Tge}
\end{align}
The quantum double Hamiltonian under a transverse field is given by
\begin{align}
    H_{\mathcal D(G)} = -\sum_v \bs A_v  - h_\text{E} \sum_e T^{1}_e - \sum_p \bs B_p 
    \label{equ:H_QD}
\end{align}
where for each vertex $v$ and plaquette $p$, we define commuting projectors
\begin{align}
    \bs A_v &= \frac{1}{|G|}\sum_{g\in G}  \left ( \prod_{e \rightarrow v}  R^g_e\right ) \left (\prod_{e \leftarrow v} L^g_e\right)\\
    \bs B_p &=\sum_{e \subset p} \delta_{1,\prod_e g_e^{O_e}}\ket{\{g_e \}}\bra{\{g_e \}} 
\end{align}
and $h_\text{E}$ is the strength of the transverse field.
Physically, $\bs A_v$ enforces a no-charge condition on the ground state, while $\bs B_p$ enforces a no-flux condition via the delta function.

We will now derive the quantum double model by starting with a lattice model of ``matter'' fields with a global $G$ symmetry, then gauging the $G$ symmetry. Here, we will opt to use a Kramers-Wannier duality to perform the gauging. In Appendix \ref{app:minimalcoupling}, we alternatively perform the gauging by minimally coupling the model to a $G$ gauge field \cite{LevinGu2012,Haegemanetal15} and show that it produces the same result.

Using the same graph as before, we will start out with a different Hilbert space. We place $G$-degrees of freedom on each vertex of the graph. The Hamiltonian is a generalization of the Ising model to $G$-variables

\begin{align}
H=-\sum_v L_v-h_\text{E}\sum_e \Delta_{e},
\label{equ:GIsingmodel}
\end{align}
where 
\begin{align}
L_v&=\frac{1}{|G|}\sum_{g\in G} L^g_v = \ket{+}\bra{+}\\
\Delta_{e}&=\sum_{g_i,g_f}\delta_{1,g_i\bar g_{f}}\ket{g_i,g_{f}}\bra{ g_i,g_{f}}.
\label{equ:Deltae}
\end{align}
Here, $\ket{+}= \frac{1}{|G|}\sum_{g \in G}\ket{g}$ is the equal-weighted superposition of all basis states.

The symmetries of this Hamiltonian are left and right group actions $\prod_v L_v^g$ and $\prod_v R^g_v$. However, throughout the rest of the paper, we will only focus on the right group action $G^R$ and its gauging.

We can see that the Hamiltonian is a transverse-field Ising model because of its two limits. When $h_\text{E}=0$, the ground state is given by a tensor product of $\ket{+}$ states and thus a paramagnet. When $h_\text{E} \rightarrow \infty$, the ground states are cat states, preferring to simultaneously point in the direction $\ket{g}$ for any $g \in G$, and therefore spontaneously breaks the global symmetry. We remark that for $G=\ZZ_N$, the Hamiltonian is simply the $\ZZ_N$ Potts model.

We will now performing a Kramers-Wannier duality, which is also known as the ``gauging map''\cite{Yoshida2017}. Heuristically, the map transforms ``$G$-spin'' variables on vertices to ``$G$-domain-wall'' variables on edges. The mapping from the vertex Hilbert space to the edge Hilbert space is that the state on each edge is determined from the two vertices it connects. Namely,
\begin{align}
   \raisebox{-.7\height}{\begin{tikzpicture}
\node[label=center:$g_{i_e} \ $] (1) at (0,0) {};
\node[label=center:$ \ g_{f_e}$] (2) at (\l,0) {};
\draw[->] (1) -- (2) node[midway, above] {};
\end{tikzpicture}} 
\rightarrow    \raisebox{-.5\height}{\begin{tikzpicture}
\node[label=center:] (1) at (0,0) {};
\node[label=center:] (2) at (\l,0) {};
\draw[->] (1) -- (2) node[midway, above] {$g_e$};
\end{tikzpicture}}
\label{equ:onsiteQR2D}
\end{align}
where $g_e =g_{i_e}\bar g_{f_e}$. One can immediately notice that the map turns the Ising term $\Delta_e$ into a transverse field $\ket{1}\bra{1}_e$ for each edge, which is the projector $T^1_e$.

Using this map, the operators and symmetries in the Hamiltonian map as
\begin{align} L^g_v &\rightarrow \bs A^g_v =   \prod_{e \rightarrow v}  R^g_e \times \prod_{e \leftarrow v} L^g_e,\\
\label{equ:Agv}
\Delta_e &\rightarrow \sum_{g_e} \delta_{1,g_e} \ket{g_e}\bra{g_e}= T^1_e,\\
\prod_v R^g_v &\rightarrow \mathbbm 1.
\end{align}
Therefore, this procedure effectively ``gauges'' the right-multiplication symmetry $G^R$ of the Hamiltonian.

Next, we notice that the map also produces constraints. For any plaquette $p$, we have the following identity
\begin{align}
    \prod_{e \in p} (g_{i_e}\bar g_{f_e})^{O(e)}=1,
\end{align}
where the product is ordered clockwise around the plaquette. This implies that one must enforce the following fluxless condition
\begin{align}
    \prod_{e \in p} g_e^{O(e)}=1
\end{align}
in the gauged Hamiltonian, which we can enforce energetically with the term $\bs B_p$. To conclude, the gauged Hamiltonian is
\begin{align}
H_\text{gauged}=-\sum_v \bs A_v-h_\text{E}\sum_e T^1_e - \sum_p \bs B_p,
\end{align}
which is precisely the model for $\mathcal D(G)$ under a transverse field introduced earlier. The gauging description here establishes that gauging a global symmetry of a $G$-paramagnet gives a $G$-topological order\footnote{We note that after gauging there is still a residual global symmetry given by $G^L/Z(G) \equiv \text{Inn}(G)$ in the quantum double model, but it is not relevant to our discussion.}. 

\subsection{ Ising model and $\mathcal D(G)$ in the factored basis}\label{app:factored}
Now, let us alternatively perform the gauging in two separate steps. First we will gauge a normal subgroup $N \triangleleft G$ to obtain a topological order with gauge group $N$ enriched by the quotient group $Q$, and second, we then gauge $Q$. As a byproduct of two separate steps, we obtain an intermediate $N$-topological order with symmetry enrichment such as permutation and fractionalization of the anyons, which is encoded in the data of the group extension. We give examples for $G=D_4$ and different choices of normal subgroups $N$ in Appendix \ref{app:SymmetryEnrichment}.

Using the factored system reviewed in section \ref{sec:factor}, we can use it to ``factor'' a $G$ d.o.f. into a $N$ d.o.f. tensored with a $Q$ d.o.f. That is, we can define a new basis $\ket{n} \otimes \ket{q} \in \CC[N] \otimes \CC[Q] \cong \CC[G]$. We will refer to this basis as the \textit{factored basis}. Let us write explicitly how the operators we defined previously are written in the factored basis. We start with left and right multiplications. To do this, we compute its action on a generic basis element $\ket{g_e}=\ket{\fs{n_e}{q_e}}$ in the factored basis.
\begin{align}
  L^{g}_e\ket{g_e}&=  \ket{\fs{n}{q}\cdot \fs{n_e}{q_e}} = \ket{\fs{\omega(q,q_e)n\aut{q}{n_e}}{qq_e}}\\
  R^{g}_e\ket{g_e}&=  \ket{\fs{n_e}{q_e} \cdot \lbar{\fs{n}{q}}} = \ket{\fs{n_e}{q_e} \cdot \fs{ \bar \omega(\bar q,q) \sigma_{\bar q}[\bar n]}{\bar q}}  \nonumber \\
  &=\ket{ \fs{\omega(q_e,\bar q)n_e\aut{q_e}{\bar \omega(\bar q ,q)\sigma_{\bar q}[\bar n]}}{q_e\bar q} }\nonumber\\
  &= \ket{\fs{n_e\bar \omega(q_e\bar q,q)\aut{q_e\bar q}{\bar n}}{q_e\bar q}}
\end{align}
Therefore, using the controlled operators, the left and right multiplication in the factored basis acts as
\begin{align}
L^{g}_e &= (\mathbbm 1 \otimes L^q_e ) \times (L_e^{\Omega(q,e)n} \Sigma^q_e \otimes \mathbbm 1)\label{equ:Lgfactored} \\
R^{g}_e &= (R_e^{\Omega(e,q)\Sigma^e[n]} \otimes \mathbbm1) \times (\mathbbm 1 \otimes R^q_e ) \label{equ:Rgfactored} 
\end{align}
Note that the ordering of operators is of importance here. 

Moving on to the Ising term in Eq. \eqref{equ:Deltae} , we can separate it into two delta functions
\begin{align}
\begin{split}
    \Delta_{e}=&\sum_{g_i,g_f}\delta_{1,g_i\bar g_{f}}\ket{g_i,g_{f}}\bra{ g_i,g_{f}}\\
    =& \sum_{g_i,g_f} \delta_{1,t(g_i\bar g_f)}\delta_{1,\pi(g_i\bar g_f)} \ket{g_i,g_{f}}\bra{ g_i,g_{f}}\\
    =&\sum_{n_i,n_f,q_i,q_f} \delta_{1,\bar \omega(q_i\bar q_f, q_f) n_i\aut{q_i\bar q_f}{\bar n_f}} \\
    &\delta_{1,q_i\bar q_f}\ket{n_i,n_f,q_i,q_f}\bra{n_i,n_f,q_i,q_f}
\end{split}
\end{align}

The terms in the quantum double Hamiltonian in Eq. \eqref{equ:H_QD} can also be rewritten in the factored basis. The vertex term $\bs A^g_v$ is the same using Eq. \eqref{equ:Agv} and the explicit form of left and right multiplications in the factored basis given above. For the plaquette term $\bs B_p$, we can separate the expression into two delta functions, one of which contains only elements in $Q$. For simplicity, we derive the plaquette term for the square lattice. For a plaquette labeled as
\begin{center}
   \raisebox{-.5\height}{\begin{tikzpicture}
\node[vertex] (1) at (0,0) {};
\node[vertex] (2) at (0,\l) {};
\node[vertex] (3) at (\l,\l) {};
\node[vertex] (4) at (\l,0) {};
\draw[->] (1) -- (2) node[midway, left] {$g_{1}$};
\draw[->] (2) -- (3) node[midway, above] {$g_{2}$};
\draw[->] (1) -- (4) node[midway, below] {$g_{4}$};
\draw[->] (4) -- (3) node[midway,right] {$g_{3}$};
\end{tikzpicture}} 
\end{center}
The plaquette term is
\begin{align}
   \bs B_p = \sum_{g_1,g_2,g_3,g_4} \delta_{1,g_1g_2\bar g_3 \bar g_4} \ket{g_1,g_2,g_3,g_4}\bra{g_1,g_2,g_3,g_4} 
\end{align}
Using the multiplication of the factor system, one finds
\begin{align}
    t(g_1g_2\bar g_3\bar g_4) =& \omega(q_1,q_2) \omega(q_1q_2,\bar q_3) \omega(q_1q_2\bar q_3,q_4) n_1\nonumber\\
    &\aut{q_1}{n_2} \aut{q_1q_2}{\bar \omega (\bar q_3,q_3)} \aut{q_1q_2\bar q_3}{\bar n_3} \nonumber \\
    &\aut{q_1q_2\bar q_3}{\bar \omega(\bar q_4,q_4)} \aut{q_1q_2\bar q_3\bar q_4}{n_4}\\
    \pi(g_1g_2\bar g_3\bar g_4) =&q_1q_2\bar q_3\bar q_4
\end{align}
Thus, we obtain two delta functions setting the above two quantities to one. In addition, if we impose $q_1q_2\bar q_3\bar q_4=1$ from the second delta function in addition to applying appropriate cocycle conditions, the first constraint simplifies to
\begin{align}
    \omega(q_1,q_2)\bar \omega (q_4,q_3)  n_1 \aut{q_1}{n_2} \aut{q_4}{\bar n_3} n_4=1 
    \label{equ:Bp_nconstraint}
\end{align}
Therefore, the plaquette term in the factored basis is
\begin{align}
   \bs B_p = \sum_{\{n_i\},\{q_i\}} &\delta_{1,\omega(q_1,q_2)\bar \omega (q_4,q_3)  n_1 \aut{q_1}{n_2} \aut{q_4}{\bar n_3} n_4}\nonumber\\
   &\delta_{1,q_1q_2\bar q_3\bar q_4} \ket{\{n_i\},\{q_i\}}\bra{\{n_i\},\{q_i\}}
\end{align}

\subsection{Gauging $N$}\label{sec:GaugeN2D}
As with gauging $G$, gauging a normal subgroup $N$ can be thought of as a Kramers-Wannier duality which maps $G$-matter variables on vertices to a model with $N$-gauge variables on edges and $Q$-matter variables on vertices. The resulting model is a Symmetry-Enriched Topological (SET) order, an $N$-topological order enriched by a global symmetry $Q$. For each edge, the map is given by
\begin{align}
   \raisebox{-.7\height}{\begin{tikzpicture}
\node[label=center:$g_i \ $] (1) at (0,0) {};
\node[label=center:$ \ g_f$] (2) at (\l,0) {};
\draw[->] (1) -- (2) node[midway, above] {};
\end{tikzpicture}} 
\rightarrow    \raisebox{-.5\height}{\begin{tikzpicture}
\node[label=center:$q_i \ $] (1) at (0,0) {};
\node[label=center:$ \ q_f$] (2) at (\l,0) {};
\draw[->] (1) -- (2) node[midway, above] {$n_e$};
\end{tikzpicture}}
\label{equ:gaugingmap2D}
\end{align}
where
\begin{align}
  q_i &= \pi(g_i), & q_f&=\pi(g_f), & n_e &= t(g_i\bar g_f).
    \label{equ:ne2D}
\end{align}
Conceptually, the map can be thought of as a projection to $Q$ variables via $\pi$ and to $N$-domain walls via $t$. To work out how the operators in the Ising model map, we use the  identities \eqref{equ:pimaponproduct}-\eqref{equ:multiply_n_in_t}. 

As an example, let us work out how $L^g_v$ maps under the duality. Under the map, the operator $L^g_v$ acts on the $Q$-d.o.f. at site $v$ as $L^q_v$, but it also now act on the $N$-d.o.f. on edges surrounding $v$. For edges $e \rightarrow v$, $L^g_v$ acts as left multiplication on $i_e$. Since such edge $e$ has the d.o.f. $n_e = t(g_i\bar g_f)$, using Eq. \eqref{equ:Lactoni}, we obtain that the action of $L^g_v$ on the dual d.o.f. sends $n_e$ to $\bar \omega(q,q_i\bar q_f) n \aut{q}{n_e}$, which can be implemented via the gate $L_e^{ \bar \Omega(q,v\bar f) n} \Sigma^q_e$. Similarly, for edges $e \leftarrow v$, we can use \eqref{equ:Lactonf} to obtain the action of $L^g_v$ on such edges. We conclude that under the duality,
\begin{align}
    L^{g}_v \rightarrow  A^g_v= \prod_{e\rightarrow v}R^{\Omega(i\bar v ,q)\Sigma^{i\bar v}[n]}_e  \times L^q_v \times\prod_{e\leftarrow v} L_e^{ \bar \Omega(q,v\bar f) n} \Sigma^q_e.
    \label{equ:AgvSET}
\end{align}

For the Ising term, it is apparent that
\begin{align}
    \Delta_e \rightarrow \tau_e =T^1_e \sum_{q_i,q_f} \delta_{1,q_i\bar q_f}\ket{q_i,q_f}\bra{q_i,q_f}
\end{align}
As for the symmetry, since $\prod_v R^g_v$ leaves $t(g_i\bar g_f)$ invariant for every edge, it only maps to right multiplication by $q$ on every vertex.
\begin{align}
    \prod_v R^g_v \rightarrow \prod_v R^q_v
\end{align}

Finally, we enforce the constraints from this mapping energetically as a fluxless condition for each plaquette. We notice that by defining
\begin{align}
 \tilde n_e = \aut{\bar q_i}{  \omega(q_i\bar q_f, q_f) n_e} =   \aut{\bar q_i}{ n_i}\aut{\bar q_f}{\bar n_f},
 \label{equ:tilden_e}
\end{align}
the product of the right hand side around a closed loop $l$ (taking care of orientation) is the identity. Therefore,
\begin{align}
    \prod_{e \in l}  \tilde n_e^{O_e} =1.
    \label{equ:Nconstraint}
\end{align}
This constraint can be enforced energetically via the following plaquette term
\begin{align}
 B^N_p =\sum_{\{n_e,q_v\}}\delta_{1,  \prod_{e \subset p}  \tilde n_e^{O_e}} \ket{\{n_e,q_v\}}\bra{\{n_e,q_v\}}.
 \label{equ:BN2D}
\end{align}
To conclude, the Hamiltonian for the SET is
\begin{align}
H_{SET} =& -\frac{1}{|G|} \sum_v \left (\sum_{g \in G} A^{g}_v \right) - h_\text{E}\sum_e \tau_e - \sum_p B^N_p
\label{equ:SET}
\end{align}

\begin{table*}[t]
    \centering
        \caption{Mapping of operators in the $G$-Ising model, the SET ($N$-topological order enriched by $Q$), and the quantum double $\mathcal D(G)$ in the factored basis. Here, the operators are defined on a square lattice for simplicity.}
    \begin{tabular}{|c|c|c|}
    \hline
   $G$-Ising & SET & $\mathcal D(G)$ \\
     \hline
 $ L^{g}_v $    &$\displaystyle A^g_v= \prod_{e\rightarrow v}R^{\Omega(i\bar v ,q)\Sigma^{i\bar v}[n]}_e  \times L^q_v \times\prod_{e\leftarrow v} L_e^{ \bar \Omega(q,v\bar f) n} \Sigma^q$ &$\displaystyle \bs A^g_v= \prod_{e \rightarrow v}  \left [(R_e^{\Omega(e,q)\Sigma^e[n]} \otimes \mathbbm1) \times (\mathbbm 1 \otimes R^q_e )\right]$\\
 && $\displaystyle \times \prod_{e \leftarrow v} \left [(\mathbbm 1 \otimes L^q_e ) \times (L_e^{\Omega(q,e)n} \Sigma^q_e \otimes \mathbbm 1)\right] $\\
  \hline
 $\Delta_e$ & $\displaystyle \tau_e =T^1_e \sum_{q_i,q_f} \delta_{1,q_i\bar q_f} \ket{q_i,q_f}\bra{q_i,q_f}$ & $\bs \tau_e = T^1_e \otimes T^1_e$ \\
  \hline
 $1$ & $\displaystyle  B^N_p =\sum_{\{n_e,q_v\}}\delta_{1,  \prod_{e \subset p}  \tilde n_e^{O_e}} \ket{\{n_e,q_v\}}\bra{\{n_e,q_v\}}$ & $ \displaystyle \bs B^N_p = \sum_{\{n_e,q_e\}} \delta_{1,  \omega(q_1,q_2)\bar \omega (q_4,q_3)  n_1 \aut{q_1}{n_2} \aut{q_4}{\bar n_3} n_4 } \ket{\{n_e,q_e\}}\bra{\{n_e,q_e\}}$ \\
  \hline
 $1$ & $1$ & $\displaystyle \bs B^Q_p = \sum_{\{q_e\}} \delta_{1, q_1 q_2 \bar q_3 \bar q_4  }\ket{\{q_e\}}\bra{\{q_e\}}$\\
  \hline
  $\displaystyle \prod_v R^g_v$ & $\displaystyle \prod_v R^{q}_v$ & $ 1$ \\
    \hline
    \end{tabular}
    \label{tab:duality2D}
\end{table*}

\subsection{Gauging $Q$ of the SET} \label{sec:gaugeQ2D}
We would now like to gauge the remaining $Q$ symmetry of the SET using the gauging map $\ket{q_i,q_f} \rightarrow \ket{q_i\bar q_f}$. For the vertex term $A^g_v$, first, we can replace $i\bar v$ and $v\bar f$ with $e$ for edges $e\rightarrow v$ and $e\leftarrow v$ in Eq. \eqref{equ:AgvSET}, respectively. In addition, we gauge the second term $L^q_v$ to $\prod_{e \rightarrow v} R^q_e \times \prod_{e \leftarrow v} L^q_e$ as usual. Keeping track of the tensor products, we obtain
\begin{align}
A^g_v \rightarrow \bs A^g_v =&\prod_{e\rightarrow v} \left( R^{\Omega(e ,q)\Sigma^{e}[n]}_e \otimes \mathbbm 1 \right)  \times\prod_{e \rightarrow v}\left(\mathbbm 1 \otimes R^q_e \right) \nonumber\\
&\times \prod_{e \leftarrow v}\left(\mathbbm 1 \otimes L^q_e\right) \times\prod_{e\leftarrow v} \left ( L_e^{ \bar \Omega(q,e) n} \Sigma^q_e \otimes \mathbbm 1 \right).
\end{align}
Finally, using Eqs.\eqref{equ:Lgfactored}-\eqref{equ:Rgfactored}, we see that the expression matches $\bs A^g_v$ in the unfactored basis \eqref{equ:Agv} as expected.

To gauge the plaquette term, $B^N_p$, we will do this on the square lattice for simplicity. We label the vertices and edges with variables as shown
\begin{center}
   \raisebox{-.5\height}{\begin{tikzpicture}
\node[label=center:$q_a$] (1) at (0,0) {};
\node[label=center:$q_b$] (2) at (0,\l) {};
\node[label=center:$q_c$] (3) at (\l,\l) {};
\node[label=center:$q_d$] (4) at (\l,0) {};
\draw[->] (1) -- (2) node[midway, left] {$n_{1}$};
\draw[->] (2) -- (3) node[midway, above] {$n_{2}$};
\draw[->] (1) -- (4) node[midway, below] {$n_{4}$};
\draw[->] (4) -- (3) node[midway,right] {$n_{3}$};
\end{tikzpicture}} 
\end{center}
Then, the constraint of the delta function of $B^N_p$ in Eq. \eqref{equ:BN2D} reads
\begin{align}
     \aut{\bar q_{i_1}}{\omega(q_{i_1}\bar q_{f_1},q_{f_1})n_1} \aut{\bar q_{i_2}}{\omega(q_{i_2}\bar q_{f_2},q_{f_2})n_2}&\nonumber\\
     \aut{ \bar q_{i_3}}{\bar \omega(q_{i_2}\bar q_{f_3},q_{f_3})\bar n_3} \aut{\bar q_{i_4}}{\bar \omega(q_{i_4}\bar q_{f_4},q_{f_4})n_4} &=1
     \label{equ:constraint2Dsquarelattice}
\end{align}
From the figure, we can relabel the sites $i_1 = i_4 \equiv a$, $f_1=i_2\equiv b$, $f_2=f_3\equiv c$, $f_4=i_3\equiv d$. We apply $\sigma_{q_{a}}$ on both sides to obtain
\begin{align}
  \omega(q_{a}\bar q_{b},q_{b}) \aut{q_a\bar q_b}{ \omega(q_{b}\bar q_{c},q_{c})} \aut{q_a\bar q_d}{\bar \omega(q_{d}\bar q_{c},q_{c})} &\nonumber\\ \bar \omega(q_{a}\bar q_{d},q_{d})  n_1 \aut{q_{a}\bar q_{b}}{n_2} \aut{q_{a}\bar q_{d}}{\bar n_3} n_4 &=1
\end{align}
Using the cocycle conditions to get rid of the cocycles with $\sigma$ acting on them, the expression simplifies to
\begin{align}
  \omega(q_{a}\bar q_{b},q_{b}\bar q_c) \bar \omega(q_{a}\bar q_{d},q_{d}\bar q_c)  n_1 \aut{q_{a}\bar q_{b}}{n_2} \aut{q_{a}\bar q_{d}}{\bar n_3} n_4 =1
  \label{equ:constraint2Dsquarelatticesimplified}
\end{align}

We can now use the gauging map
\begin{align}
    q_{a}\bar q_b &\rightarrow q_1 &   q_{b}\bar q_c &\rightarrow q_2&  q_{d}\bar q_c &\rightarrow q_3&  q_{a}\bar q_d &\rightarrow q_4 
\end{align}
to map the above constraint to
\begin{align}
\omega(q_1,q_2)\bar \omega (q_4,q_3)  n_1 \aut{q_1}{n_2} \aut{q_4}{\bar n_3} n_4  =1
\end{align}
which is exactly the constraint that appears in Eq. \eqref{equ:Bp_nconstraint}.  The plaquette term of the SET therefore gauges to
\begin{align}
  B^N_p \rightarrow  \bs B^N_p = \sum_{\{n_e,q_e\}}& \delta_{1,  \omega(q_1,q_2)\bar \omega (q_4,q_3)  n_1 \aut{q_1}{n_2} \aut{q_4}{\bar n_3} n_4 }\nonumber \\
  &\ket{\{n_e,q_e\}}\bra{\{n_e,q_e\}}
\end{align}
In addition, we also need to enforce a fluxless condition for the new $Q$ variables on each plaquette. This is done via
\begin{align}
    \bs B^Q_p =  \sum_{\{q_e\}} \delta_{1, q_1 q_2 \bar q_3 \bar q_4  }\ket{\{q_e\}}\bra{\{q_e\}}
\end{align}
Therefore, after gauging $Q$ of the SET, we obtain
\begin{align}
 H_\text{gauged SET} =& -\sum_v \bs A_v  - h_\text{E} \sum_e (T^{1}_e \otimes T^1_e)\nonumber \\ &- \sum_p \left(\bs B^N_p  + \bs B^Q_p\right)
    \label{equ:Ham_gaugedSET}
\end{align}
It is important to note that the Hamiltonian above is not exactly the quantum double Hamiltonian (under a transverse field) defined in Eq. \eqref{equ:H_QD}. To be precise, although the $\bs A_v$ terms match exactly, the plaquette terms $\bs B_p$ differ. The reason is because by performing two separate gaugings, we have imposed fluxless conditions on two occasions, which resulted in two separated plaquette terms $\bs B^N_p$ and $\bs B^Q_p$ rather than a single plaquette term $\bs B_p$ of the quantum double model. Nevertheless, since the two plaquette terms are commuting projectors and $\bs B^N_p\bs B^Q_p = \bs B_p$, the two Hamiltonians have the same ground state, and therefore describe the same topological order.

To summarize the dualities, we list the result of operators under the gauging map, first by gauging $N$ followed by gauging $Q$ in Table \ref{tab:duality2D}.

\section{Gauging $G$ via minimal coupling}\label{app:minimalcoupling}
In this appendix, we use the minimal coupling procedure instead of the Kramers-Wannier duality to demonstrate the mapping from a $G$-Ising model to a $G$-topological order under a transverse field.\cite{LevinGu2012,Haegemanetal15}

The minimal coupling procedure we will perform  promotes the right multiplication  symmetry $G^R$ into a local symmetry. We add $G$ gauge fields living on each edge and impose the following Gauss law on each vertex
\begin{align}
   \mathcal G_v &= \frac{1}{|G|}\sum_g \mathcal G^g_{v} =1\\
 \mathcal G^g_{v} &= \left (\prod_{e \rightarrow v}R^g_e \right) R^g_v \left (\prod_{e \leftarrow v}L^g_e \right)
\end{align}
It is apparent that $\mathcal G^g_{v}$ on different vertices commute. Therefore, the Gauss law on different vertices commute. Furthermore, $\mathcal G^g_{v}$ obeys the group law of $G$, and they therefore implement a local symmetry transformation. Multiplying $\mathcal G^g_{v}$ on all vertices, we obtain
\begin{align}
    \prod_v R^g_v \prod_e (L^g_eR^g_e),
\end{align}
which, after further removing the conjugation symmetry on the edge (gauge) variables, recovers the global $G^R$ symmetry.

Next, we must minimally couple the Hamiltonian so that it commutes with the Gauss law.  The minimally coupled term is
\begin{align} 
    \Delta_{e}^\text{MC}&=\sum_{g_i,g_e,g_f}\delta_{1,g_ig_{e}g_{f}^{-1}}\ket{g_i,g_{e},g_{f}}\bra{g_i,g_{e},g_{f}}.
\end{align}
One can see that it commutes with each $\mathcal G^g_{v}$, and therefore commutes with $\mathcal G_v$. The projector $\ket{+}\bra {+}$ already commutes with the Gauss law and so it does not need to be modified. Finally we enforce a ``fluxless'' condition around each plaquette $p$. Similarly, we enforce this energetically using $\bs B_p$ defined in the main text. The minimally coupled Hamiltonian reads
\begin{align}
H^\text{MC}=-\sum_e \Delta_{e}^\text{MC} -h\sum_v L_v - \Lambda \sum_p \bs B_p; \ \ \ \ \ \ \mathcal G_v =1
\end{align}

Next, we will compute the effective Hamiltonian obtained by restricting the Hilbert space using the Gauss law, and show that it is the quantum double model under a transverse field. To do so, we will perform a change of basis using a unitary $U$ such that the Gauss law becomes
\begin{align}
   U\mathcal  G_v U^{\dagger}=\frac{1}{|G|}\sum_{g \in G} R^g_v =1
\end{align}

The unitary is given by
\begin{align}
   U=\prod_v\left [ \prod_{e\rightarrow v} CR_{ve} \prod_{e\leftarrow v} CL_{ve}  \right]
\end{align}
where
\begin{align}
CL_{ve}: \ket{a}_v\ket{b}_e &\rightarrow \ket{a}_v\ket{ab}_e\\
CR_{ve}: \ket{a}_v\ket{b}_e &\rightarrow \ket{a}_v\ket{b\bar a}_e
\end{align}
are controlled left and right multiplication operators\cite{Brell2015}, which satisfy the following identities
\begin{align}
CL_{ve}(L^g_v \otimes \mathbbm1_e)CL^{-1}_{ve} &= L^g_v \otimes L^g_e,\\
CR_{ve}(L^g_v \otimes \mathbbm1_e)CL^{-1}_{ve} &= L^g_v \otimes R^g_e,\\
CL_{ve}(R^g_v \otimes L^g_e)CL^{-1}_{ve} &= R^g_v \otimes  \mathbbm 1_e,\\
CR_{ve}(R^g_v \otimes R^g_e)CR^{-1}_{ve} &= R^g_v \otimes  \mathbbm 1_e.
\end{align}
The minimally coupled kinetic term in the new basis is
\begin{align}
    U \Delta^\text{MC} U^\dagger&=  \sum_{g_i,g_e,g_f} U \Delta^\text{MC} \ket{g_i,\bar g_ig_eg_f ,g_f} \bra{g_i,g_e,g_f}\nonumber \\
  &= \sum_{g_i,g_e,g_f} U \delta_{1,g_i(\bar g_ig_eg_f)\bar g_f}\ket{g_i,\bar g_ig_eg_f ,g_f}\bra{g_i,g_e,g_f} \nonumber\\
  &= \sum_{g_i,g_e,g_f} \delta_{1,g_e} \ket{g_i,g_e,g_f}\bra{g_i,g_e,g_f} =\ket{1}\bra{1}_e
\end{align}
A similar calculation shows that $\bs B_p$ is invariant under $U$. Therefore,
\begin{align}
UH^\text{MC}U^{\dagger}=&-\sum_v \left [ \frac{1}{|G|}\sum_{g\in G}\left ( \prod_{e \rightarrow v}  R^g_e\right ) L^g_v \left (\prod_{e \leftarrow v} L^g_e\right) \right] \nonumber \\
&  -h_\text{Ising}\sum_e\ket{1}\bra{1}_e - \sum_p \bs B_p
\end{align}
with Gauss law  $U\mathcal G_vU^{\dagger} = \frac{1}{|G|} \sum R^g_v =1$

Now, we restrict to the subspace invariant under the Gauss law, which is the singlet $\ket{+}_v$ for every vertex. This means that $L^g_v$ acts as the trivial irrep in this subspace so we can replace $L^g_v =1$ for all $g$. The result is the effective Hamiltonian

\begin{align}
H_\text{gauge}=&-\sum_v \left [ \frac{1}{|G|}\sum_{g\in G}\left ( \prod_{e \rightarrow v}  R^g_e\right ) \left (\prod_{e \leftarrow v} L^g_e\right) \right] \nonumber\\
&-h_\text{Ising}\sum_e \ket{1}\bra{1}_e -\sum_p \bs B_p
\end{align}
This is the quantum double model under a transverse field.

\section{Examples of Symmetry Enrichment in Topological and Fracton Phases}\label{app:SymmetryEnrichment}

In this Appendix, we give examples of symmetry fractionalization in topological and fracton phases. We emphasize that the phenomena of the former has already been established both mathematically\cite{BarkeshliBondersonChengWang2019} and explicitly via lattice models\cite{Hermele2014,TarantinoLinderFidkowski2016}. Various other constructions including those that perform a general permutation and fractionalization have also been demonstrated  \cite{MesarosRan2013,Ben-ZionDasMcGreevy16,HeinrichBurnellFidkowskiLevin22016,ChengGuJiangQi2017,WilliamsonBultinckVerstraete17,Garre-RubioIblisdir19}. Nevertheless, we find it insightful to demonstrate the enrichment properties explicitly for the SETs we present in Appendix \ref{sec:GaugeN2D} since it is generalizes naturally to the fracton case where a global symmetry permutes or fractionalizes the fractonic excitations as presented in the main text.
\subsection{Permutation and Fractionalization in the SET}\label{app:GaugingN}

Starting with the SET Hamiltonian \eqref{equ:SET}, we will perform a basis transformation to turn the $N$-d.o.f. on edges from $n_e$ to $\tilde n_e$ defined in Eq. \eqref{equ:tilden_e}. In this basis, the d.o.f. on $N$ forms the quantum double model $D(N)$, which can be partially seen from the expression of the plaquette term $B^N_p$ \eqref{equ:BN2D}. We can then explicitly demonstrate the enrichment via the symmetry $Q$ via the fractionalization and permutation of anyons. 

There are two ways in which this basis transformation can be achieved. The first is to conjugate the Hamiltonian by the following change of basis unitary
\begin{align}
    U = \prod_e \Sigma^i_e L^{\Omega(i\bar f,f)}_e.
    \label{equ:SETchangeofbasis}
\end{align}
The second is to use the following gauging map instead of Eq. \eqref{equ:gaugingmap2D} to dualize the $G$-Ising model.
\begin{align}
   \raisebox{-.7\height}{\begin{tikzpicture}
\node[label=center:$g_i \ $] (1) at (0,0) {};
\node[label=center:$ \ g_f$] (2) at (\l,0) {};
\draw[->] (1) -- (2) node[midway, above] {};
\end{tikzpicture}} 
\rightarrow    \raisebox{-.5\height}{\begin{tikzpicture}
\node[label=center:$q_i \ $] (1) at (0,0) {};
\node[label=center:$ \ q_f$] (2) at (\l,0) {};
\draw[->] (1) -- (2) node[midway, above] {$\tilde n_e$};
\end{tikzpicture}}
\end{align}
with $\tilde n_e$ defined as Eq. \eqref{equ:tilden_e}.

The two methods are equivalent. Here we will choose to present the second method to calculate the form of the operators, which we will note with tildes to contrast them from the operators in the basis of Eq. \eqref{equ:SET} presented earlier. That is, all terms with tildes are related to the ones without tildes via a conjugation of $U$.

First, we will need to derive how the expression $\aut{q_v}{n_v}$ transforms under left and right multiplication. They are
\begin{align}
    L^g: \aut{q_v}{n_v} &\rightarrow \aut{\overline{qq_v}}{\omega(q,q_v)n \aut{q}{n_v}} \label{equ:Lg_On_sigma}\\
    L^g: \aut{q_v}{\bar n_v} &\rightarrow \aut{\overline{qq_v}}{\overline{\omega(q,q_v)n \aut{q}{n_v}}}\\
    R^{\bar g}:\aut{q_v}{n_v} &\rightarrow  \aut{q \bar q_v}{ \omega(q_v,q) n_v\aut{q_v}{n}}\\
    R^{\bar g}:\aut{q_v}{n_v} &\rightarrow  \aut{q \bar q_v}{\overline{ \omega(q_v,q) n_v\aut{q_v}{n}}}
    \label{equ:Rg_On_sigmainv}
\end{align}

Using this, we can figure out how the global symmetry $\prod_v R^{\bar g}_v$ maps. In addition to acting as $R^{\bar q}_v$ on all vertices, it also has a non-trivial action on $\tilde n_e$
\begin{align}
  \prod_v R^{\bar g}_v: &\tilde n_e  =\aut{\bar q_i}{n_i}\aut{\bar q_f}{\bar n_f} \nonumber\\
  &\rightarrow  \aut{q \bar q_i}{ \omega(q_i,q) n_i\aut{q_i}{n}} \aut{q \bar q_f}{\overline{ \omega(q_f,q) n_f\aut{q_f}{n}}}\nonumber\\
  &=\aut{q} { \aut{\bar q_i}{\omega(q_i,q)}\aut{\bar q_f}{\bar \omega(q_f,q)}\tilde{n}_e}
\end{align}

Therefore, the global $G^R$ symmetry dualizes to
\begin{align}
    \prod_v R^{\bar g}_v \rightarrow \prod_v R^{\bar q}_v \times \prod_e \Sigma^{ q}_e \times \prod_e  L^{\Sigma^{\bar i}_e[\Omega_e(i, q)] \Sigma^{\bar f}_e[\bar \Omega_e(f, q)]}_e.
\end{align}
So in this basis, the remaining $Q^R$ symmetry is no longer onsite when $\omega$ is a non-trivial cocycle.

Next, let us map $L^g_v$. For edges $e\rightarrow v$, $v=i_e$, so $\tilde n_e$ transforms as 
\begin{align}
  L^g_{v}: \tilde n_e \rightarrow&  \aut{\overline{qq_v}}{\omega(q,q_v)n \aut{q}{n_v}}\aut{\bar q_f}{\bar n_f} \nonumber\\
  &= \aut{\overline{qq_v}}{\omega(q,q_v)n}\tilde{n}_{e}
\end{align}
And for edges $e\leftarrow v$, $v=f_e$, so $\tilde n_e$ transforms as 
\begin{align}
  L^g_{v}: \tilde n_e \rightarrow&  \aut{\bar q_i} {n_i}\aut{\bar q_v\bar q}{\lbar {\omega(q,q_v)n \aut{q}{n_v}}} \nonumber\\
  &= \aut{\overline{qq_v}}{\bar \omega(q,q_v)\bar n}\tilde{n}_e
\end{align}

Therefore, we see that the action on the dual variables is
\begin{align}
  L^g_v \rightarrow \tilde A^g_v=\prod_{e\rightarrow v}R^{\Sigma^{\bar v}[ \Omega(q,\bar qq_v) n]}_{e}     \times\prod_{e\leftarrow v} L^{ \Sigma^{\bar v}[\Omega(q,\bar qq_v)n]}_e \times L^q_v
\end{align}
In particular, we see that for elements $\fs{n}{1}$ in the embedded normal subgroup
\begin{align}
\tilde A^{\fs{n}{1}}_v &= \prod_{e \rightarrow v} R_e^{\Sigma^{\bar v}[n]}  \times \prod_{e \leftarrow v}L^{\Sigma^{\bar v}[n]}_e.
\end{align}
This is precisely the vertex term of $\mathcal D(N)$ up to an action by $\sigma$, which we will later demonstrate is responsible for permutation of anyons.

The constraints from the mapping is still given by Eq. \eqref{equ:Nconstraint}. However, since we are already in the $\tilde n_e$ basis, the form of the operator we enforce is
\begin{align}
    \tilde B^N_p = \sum_{\{\tilde n_e}\delta_{1,  \prod_{e \subset p}  \tilde n_e^{O_e}} \ket{\{\tilde n_e\}}\bra{\{\tilde n_e\}}.
\end{align}
This is just the plaquette term of $\mathcal D(N)$. To conclude, we have performed a basis transformation into a basis where the $N$-d.o.f. form the quantum double model $\mathcal D(N)$

\subsubsection{Examples: $G=D_4$}
To give explicit examples of symmetry fractionalization and permutation, we will analyze three examples of $G=D_4= \langle r,s | r^4 =s^2 =(sr)^2=1 \rangle$ with different choices of $N$. A stack of the examples here are in fact the realizations of the symmetry enrichment of the 1-foliated examples discussed in Sec. \ref{sec:D4examples} after gauging $N$.

1. $N= \ZZ_2 = \langle r^2| (r^2)^2 =1 \rangle$ and $Q=\ZZ_2^2 = \inner{r,s|r^2=s^2=(rs)^2=1}$ are representative elements of each coset. The extension is a central extension with cocycle specified by
\begin{align}
    \omega(r,r)&=r^2, &\omega(s,s)= \omega(rs,rs)&=1.
\end{align}
Note that the action of the cocycle on the other group elements are fixed by consistency using the cocycle condition. The vertex operators of interest are
\begin{align}
 \tilde  A^{r^2}_v &= \raisebox{-.5\height}{\includegraphics[scale=1]{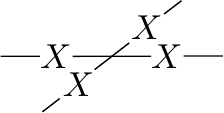}}\\
\tilde  A^{rs}_v &=
 (XI)_v \left (\raisebox{-.5\height}{\includegraphics[scale=1]{AvTC.pdf}} \right)^\frac{1-(IZ)_v}{2}\\
\tilde  A^{s}_v &=
 (IX)_v \\
 \tilde  A^{r}_v &=
 (XX)_v \left (\raisebox{-.5\height}{\includegraphics[scale=1]{AvTC.pdf}} \right)^\frac{1+(IZ)_v}{2}
\end{align}
where we have suppressed tensor products for simplicity. The above vertex operators generate all other $\tilde A^g_v$ operators. As for the plaquette term, we have
\begin{align}
    \tilde B^N_p = \frac{1+ \tilde B_p^{r^2}}{2}
\end{align}
where
\begin{align}
  \tilde B_p^{r^2} &= \raisebox{-.5\height}{\includegraphics[scale=1]{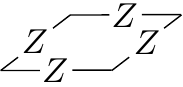}}
\end{align}

The global $\ZZ_2^2$ symmetry is generated by $\prod_v \tilde S^{rs}_v$ and $\prod_v \tilde S^{s}_v$ where
\begin{align}
 \tilde S^{rs}_v &= (XI)_v\\
    \tilde S^{s}_v&= (IX)_v \times (\tilde A^{r^2}_v)^{\frac{1-(ZI)_v}{2}}
\end{align}
This global symmetry squares to one and commutes with the terms above. Now, let us now compute the local symmetry action on a single vertex $v$. We find $(\tilde S^{rs}_v)^2 = (\tilde S^{s}_v)^2=1$, but 
\begin{align}
    (\tilde S^{r}_v)^2 = (\tilde S^{rs}_v\tilde S^{s}_v)^2 = \tilde A^{r^2}_v
\end{align}
Therefore, for an excited state where an anyon $e$ is positioned at vertex $v$, we have $\tilde A^{r^2}_v=-1$, which gives the desired fractionalization 
\begin{align}
    (\tilde S^{r}_v)^2 = -1.
\end{align}

2. $N= \ZZ_2^2 = \langle r^2,rs|(r^2)^2= (rs)^2 = (r^3s)^2=1 \rangle$ and $Q=\ZZ_2=\{r|r^2=1\}$. The extension is instead a split extension ($G= N \rtimes Q$), so $\omega(q,q') =1$. $\sigma_{r}$ acts as
\begin{align}   
\sigma_{r}[rs] &= \bar rs  &\sigma_{rs}[\bar rs] &= rs &\sigma_{r}[r^2]=r^2
\end{align}
Therefore, it is convenient to choose the Pauli operators to be in the basis generated by $rs$ and $\bar r s$, so that the global $\ZZ_2$ acts as a swap on the two copies of the toric code. In this basis, the operators are

\begin{align}
\tilde A^{rs}_v &= 
       \left (
      \raisebox{-.5\height}{\includegraphics[scale=1]{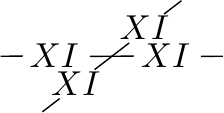}}
 \right)^\frac{1+Z_v}{2}        \left (\raisebox{-.5\height}{\includegraphics[scale=1]{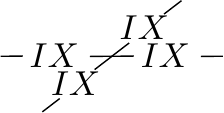}} \right)^\frac{1-Z_v}{2} \\
\tilde A^{\bar rs}_v &=\left (
      \raisebox{-.5\height}{\includegraphics[scale=1]{AvTCXI.pdf}}
 \right)^\frac{1-Z_v}{2}        \left (\raisebox{-.5\height}{\includegraphics[scale=1]{AvTCIX.pdf}} \right)^\frac{1+Z_v}{2} \\
  \tilde A_v^{r} &= X_v 
  \end{align}
  The flux term can be decomposed as
  \begin{align}
       \tilde B^N_p = \frac{1+ \tilde B_p^{rs}}{2} \cdot\frac{1+ \tilde B_p^{\bar rs}}{2}
  \end{align}
  where
  \begin{align}
 \tilde B_p^{rs} &=
    \raisebox{-.5\height}{\includegraphics[scale=1]{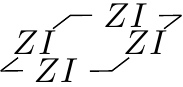}}\\
 \tilde B_p^{\bar rs} &=
     \raisebox{-.5\height}{\includegraphics[scale=1]{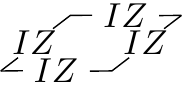}}
\end{align}
and the $\ZZ_2$ symmetry acts as $\prod_v X_v  \cdot \prod_e \textsc{swap}_e$ where
\begin{align}
\textsc{swap} &= \begin{pmatrix} 1&0&0&0\\
0&0&1&0\\
0&1&0&0\\
0&0&0&1
\end{pmatrix}.
\end{align}

A pair of pure charge and pure fluxes in the bilayer toric code labeled $e_1$, $e_2$, $m_1$ and $m_2$ are created using strings of $XI$, $IX$, $ZI$, and $IZ$ respectively. Therefore, we see that under the $\ZZ_2$ symmetry the string operators are swapped, permuting the anyons at the ends of the string.

3. $N= \ZZ_4 = \langle r|r^4=1 \rangle$ and $Q=\ZZ_2=\inner{s|s^2=1}$. The extension is also a split extension, so $\omega(q,q') =1$ and $\sigma_s$ acts as charge conjugation $\cC$ on the $N$ variables. We will use Pauli matrices $Z,X$ on $Q$ variables, $\ZZ_4$ clock and shift matrices $\cZ,\cX$ on $N$ variables, and use bars for the Hermitian conjugate (e.g. $\cC \cX \cC = \bar \cX$). The operators are
\begin{align}
  \tilde A^r_v &=
       \left (\raisebox{-.5\height}{\includegraphics[scale=1]{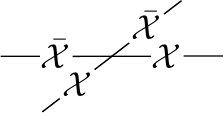}} \right)^{Z_v} \\
\tilde   A_v^s &= X_v 
\end{align}
The flux term can be written as
\begin{align}
   B^N_p = \frac{1+B_p^r + (B_p^r)^2 + (B_p^r)^3}{4} 
\end{align}
where
\begin{align}
\tilde  B_p^r &=
     \raisebox{-.5\height}{\includegraphics[scale=1]{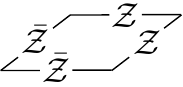}}
\end{align}
and the $\ZZ_2$ symmetry acts as $\prod_v X_v  \prod_e \cC_e$, where
\begin{align}
\cC &= \begin{pmatrix} 1&0&0&0\\
0&0&0&1\\
0&0&1&0\\
0&1&0&0
\end{pmatrix}
\end{align} is the charge conjugation operator. Thus, we have written the model in the basis of the $\ZZ_4$ toric code, with gauge charges generated by $e$ and gauge fluxes generated by $m$. Here, an $e - \bar e$ pair is created by a string of $\cZ$, and an $m-\bar m$ pair is excited with a string of $\cX$ operators on the dual lattice. The strings are charge conjugated under the global $\ZZ_2$ symmetry, which means the the symmetry permutes both $e$ and $m$ the anyons according to charge conjugation.

\begin{table*}[t!]
    \caption{Examples of possible hybrid fracton models constructed from different choices of groups $N$ and $Q$. Here, the subsystem symmetry is chosen to be three intersecting planes (3-foliated). A hybrid model (blue) is obtained for any $G$ that is a non-trivial group extension of $Q$ by $N$.  References are included for models that have appeared in previous works. The $(G,N)$ gauge theories give a unified perspective of these models. }
        \begin{tabular}{|c|c|c| c|}
    \hline
    $N$ & $Q$ & $G$ & Description\\
    \hline
    $\ZZ_1$ & $\ZZ_1$ & $\ZZ_1$ & Trivial\\
    \hline
    $\ZZ_1$ & $\ZZ_2$ & $\ZZ_2$ & $\ZZ_2$ TC\\
    \hline
    $\ZZ_2$ & $\ZZ_1$ & $\ZZ_2$ & $\ZZ_2$ X-cube\\
    \hline
  \multirow{2}{*}{$\ZZ_2$} &  \multirow{2}{*}{$\ZZ_2$} & $\ZZ_2^2$ & $\ZZ_2$ TC $\otimes$ $\ZZ_2$ X-cube\\
     &  & $\ZZ_4$ & {\color{blue}Fractonic Hybrid X-cube}\cite{TJV1}\\
     \hline
 \multirow{4}{*}{$\ZZ_2$} &  \multirow{4}{*}{$\ZZ_2^2$} & $\ZZ_2^3$ & $\ZZ_2^2$ X-cube $\otimes$ $\ZZ_2$ TC\\
&  & $\ZZ_4 \times \ZZ_2$ & {\color{blue}Fractonic Hybrid X-cube $\otimes$ $\ZZ_2$ TC}\\
 &  & $D_4$ & {\color{blue}$D_4$ hybrid model (all abelian charges mobile)}\cite{AasenBulmashPremSlagleWilliamson20}\\
  &  & $Q_8$ & {\color{blue}$Q_8$ hybrid model (all abelian charges mobile)}\\
 \hline
  \multirow{3}{*}{$\ZZ_2^2$} &  \multirow{3}{*}{$\ZZ_2$} & $\ZZ_2^3$ & $\ZZ_2^2$ X-cube $\otimes$ $\ZZ_2$ TC\\
&  & $\ZZ_4 \times \ZZ_2$ & {\color{blue}Fractonic Hybrid X-cube $\otimes$ $\ZZ_2$ X-cube}\\
 &  & $D_4$ & {\color{blue}$D_4$ hybrid model ($\bs r$ mobile)}\cite{BulmashBarkeshli2019,PremWilliamson2019,StephenGarre-RubioDuaWilliamson2020,TuChang21}\\
  \hline
      \multirow{2}{*}{$\ZZ_3$} &  \multirow{2}{*}{$\ZZ_2$} & $\ZZ_3 \times \ZZ_2$ & $\ZZ_3$ X-cube $\otimes$ $\ZZ_2$ TC\\
&  & $S_3$ & {\color{blue}$S_3$ hybrid model}\cite{TuChang21}\\
 \hline
    \multirow{4}{*}{$\ZZ_4$} &  \multirow{4}{*}{$\ZZ_2$} & $\ZZ_4 \times \ZZ_2$ & $\ZZ_4$ X-cube $\otimes$ $\ZZ_2$ TC\\
    &  & $\ZZ_8$ & {\color{blue}$\ZZ_8$ hybrid model (one mobile charge)}\\
&  & $D_4$ & {\color{blue}$D_4$ hybrid model ($\bs s$ mobile)} \\
 &  & $Q_8$ & {\color{blue}$Q_8$ hybrid model (one mobile abelian charge)}\\
    
    \hline
    \end{tabular}
    \label{tab:examples}
\end{table*}
\subsection{ Permutation and Fractionalization in the SEF}\label{app:GaugingNFracton}

Identically to the previous section, we will perform a basis transformation on the SEF Hamiltonian \eqref{equ:SEF} on the $N$-d.o.f. of each plaquette from $n_p$ to $\tilde n_p$ defined in Eq. \eqref{equ:tilden_p} reproduced here
\begin{align}
 \tilde n_p&= \aut{\bar q_i} { \omega(q_i\bar q_j,q_j)\bar \omega(q_i\bar q_k,q_k) \omega(q_i\bar q_l,q_l)n_p }\nonumber\\
 &=\aut{\bar q_i} {n_i}\aut{\bar q_j}{\bar n_j}\aut{ \bar q_k}{n_k}\aut{\bar q_l}{\bar n_l},
 \label{equ:tilden_pappendix}
\end{align}
One way to do this is to conjugate the Hamiltonian by the following change of basis unitary
\begin{align}
    U=\prod_p \Sigma^i_p L_e^{\Omega_e(i\bar j,j) \bar \Omega_e(i\bar k,k) \Omega_e(i\bar l,l)}
    \label{equ:SEFchangeofbasis}
\end{align}
The second is to use the following gauging map instead of Eq.\eqref{equ:gaugingmap3D} to dualize the $(G,N)$-Ising model.
\begin{equation}
   \raisebox{-.5\height}{\begin{tikzpicture}
\node[label=center:$g_i$] (1) at (0,0) {};
\node[label=center:$g_l$] (2) at (0,\l) {};
\node[label=center:$g_k$] (3) at (\l,\l) {};
\node[label=center:$g_j$] (4) at (\l,0) {};
\draw (1) -- (2){};
\draw (2) -- (3){};
\draw (1) -- (4) {};
\draw (4) -- (3){};
\end{tikzpicture}} \rightarrow
   \raisebox{-.5\height}{\begin{tikzpicture}
\node[label=center:$q_i$] (1) at (0,0) {};
\node[label=center:$q_l$] (2) at (0,\l) {};
\node[label=center:$q_k$] (3) at (\l,\l) {};
\node[label=center:$q_j$] (4) at (\l,0) {};
\draw[] (1) -- (2){};
\draw[] (2) -- (3){};
\draw[] (1) -- (4) {};
\draw[] (4) -- (3){};
\node[label=center:$\tilde n_p$]  at (\l/2,\l/2) {};
\end{tikzpicture}} 
\end{equation}
In an identical calculation to Appendix \ref{app:GaugingN}, we can calculate how the operators and symmetries act under this map using Eqs. \eqref{equ:Lg_On_sigma}-\eqref{equ:Rg_On_sigmainv}. The results are
\begin{align}
    \prod_v R^{\bar g}_v \rightarrow &\prod_v R^{\bar q}_v \times \prod_p  \Sigma^{\bar q}_p \\
    &\times \prod_p  L^{\Sigma^i[\Omega(i, q)] \Sigma^j[\bar \Omega(j, q)] \Sigma^k_p[ \Omega(k, q)] \Sigma^f[\bar \Omega(l, q)] }_p\nonumber\\
       L^g_v \rightarrow& \tilde A^g_v= \prod_{p|v=j,l}R^{\Sigma^{\bar v}[ \Omega(q,\bar qq_v) n]}_{p} \nonumber \\
       &\times\prod_{p|v=i,k} L^{ \Sigma^{\bar v}[\Omega(q,\bar qq_v)n]}_p \times L^q_v
\end{align}
with constraints
\begin{align}
    \tilde B^N_{c,x} =\sum_{\{\tilde n\}} \delta_{1,    \tilde n_{y}\overline{\tilde n_{z}} \ \overline{\tilde n_{\bar y}} n_{\bar z}} \ket{\{\tilde n\}}\bra{\{\tilde n\}}
    \label{equ:3DfluxXcubebasis}
\end{align}
This, along with its $C_3$ rotations are the cube terms of the X-cube model with gauge group $N$. Furthermore, for $g=\fs{n}{1}$, we see that
\begin{align}
      \tilde A^{\fs{n}{1}}_v =\prod_{p|v=j,l} R_p^{\Sigma^{\bar v}[n]} \prod_{p|v=i,k}L^{\Sigma^{\bar v}[n]}_p
\end{align}
is the vertex term of the X-cube model up to a possible action by $\sigma$.
The symmetry permutation and fractionalization of the fractons/lineons can be analogously shown for the three choices of normal subgroups in $D_4$ by replacing the the $\tilde A_v$ and $\tilde B_p$ of the toric code in 2D with $\tilde A_v$ and $\tilde B_{c,r}$ of the X-cube model in 3D.

\section{Examples of 3-foliated hybrid models}\label{app:3foliatedexamples}

In this Appendix, we give examples of 3-foliated hybrid models by specifying the groups in our commuting projector Hamiltonian \eqref{equ:GNHam}. In Table \ref{tab:examples}, we enumerate the hybrid fracton models that can be constructed given abelian groups $N$ and $Q$, many of which hosts non-abelian fractons.

\subsection{$(\ZZ_4,\ZZ_2)$}
As a sanity check, we show that the resulting model for $(\ZZ_4,\ZZ_2)$ has the same ground state as the fractonic hybrid X-cube model presented in Ref. \onlinecite{TJV1}.

We represent $N=\inner{r|r^2=1}$ and $Q=\inner{s|s^2=1}$. The group extension is a central extension whose factor system is given by $\sigma=1$ and non-trivial cocycle $\omega(s,s)=r$.

Starting with $\bs A^g_v$, the expressions reduce to the Pauli matrices as follows
\begin{align}
    R_p^r = L_p^r &\equiv X_p\\
    R_e^s = L_e^s &\equiv X_e\\
    L_p^{\Omega(e,s)}=R_p^{\Omega(e,s)} &\equiv X_p^{ \frac{1-Z_e}{2}}
\end{align}
With this, we find
\begin{align}
    \bs A^{\fs{1}{s}}_v =&  \prod_{p|v=j,k,l} X_p^{\frac{1-Z_{iv}}{2}} \times \prod_{e \supset v} X_e \nonumber \\
&\times \prod_{p|v=i} X_p^{\frac{1-Z_{vj}}{2} +\frac{1-Z_{vk}}{2} + \frac{1-Z_{vl}}{2}},\\
\bs A^{\fs{r}{1}}_v=&\prod_{p \supset v} X_p,
\end{align}
which matches Eqs. (B25)-(B26) of Ref. \onlinecite{TJV1}.

Next, we consider the $\bs B$ terms. The $\bs B^Q_{\nablapic}$ is a projector that enforces $q_{ad}q_{du}\bar q_{au}$ around a triangle $(adu)$ to be one. This is precisely the operator
\begin{align}
    \frac{1+ Z_{ad}Z_{du}Z_{au}}{2} = \frac{1 + \prod_{e \subset \nablapic}Z_e}{2}
\end{align}

 Lastly, we consider $\bs B^N_{c,x}$, which enforces 
\begin{align}
  \omega(q_{ad},q_{ds}) \omega(q_{ad},q_{dw})\omega(q_{ad},q_{du}) &\\
  \omega(q_{ac},q_{ct})\omega(q_{ac},q_{cw})\omega(q_{ac},q_{cu}) n_y n_z n_{\bar y} n_{\bar z}&=1  
\end{align}
where we have removed the bars because $N=\ZZ_2$. Define
\begin{align}
    \bs  B^{r}_{c,x} =& CZ_{ad,ds} CZ_{ad,dw} CZ_{ad,du} CZ_{ac,ct} CZ_{ac,cw} CZ_{ac,cu} \nonumber \\
     &Z_y Z_{z} Z_{\bar y} Z_{\bar z}
\end{align}
where $CZ = \text{diag}(1,1,1,-1)$ is the controlled-Z operator. The operator above satisfies the above constraint exactly when its eigenvalue is one. Therefore, we see that the projector can be written as
\begin{align}
 \bs B^N_{c,x} = \frac{1+ \bs B^{r}_{c,x}}{2}.   
\end{align}
In Ref. \onlinecite{TJV1}, we instead have the projector $\frac{1 + \bs B_{c,x}+ \bs B_{c,x}^2 + \bs B_{c,x}^3}{4}$ where
\begin{align}
    \bs B_{c,x} = & S_{ad} S_{at} S_{ds}  S_{du}  S_{cw} S^\dagger_{ac} S^\dagger_{as} S^\dagger_{dw} S^\dagger_{cu} S^\dagger_{ct}    \nonumber \\
     &Z_y Z_{z} Z_{\bar y} Z_{\bar z}
\end{align}
These two projectors are not equal. Physically, this came from the fact that we started with different plaquette terms that commuted with the $(\ZZ_4,\ZZ_2)$ symmetry in the paramagnet. Nevertheless, we can verify via an explicit calculation that
\begin{align}
    \bs B^N_{c,x} \prod_{\nablapic \subset c} \bs B_{\nablapic}^Q = \frac{1 + \bs B_{c,x}+ \bs B_{c,x}^2 + \bs B_{c,x}^3}{4} \prod_{\nablapic \subset c} \bs B_{\nablapic}^Q
\end{align}
That is, the projectors are equivalent in the $\bs B^Q_{\nablapic}=1$ subspace. Therefore, we conclude that although the Hamiltonians are different, the two models have exactly the same ground state.

\subsection{$(S_3,\ZZ_3)$}
Next, we present the simplest non-abelian hybrid fracton order, whose charges transform as irreps of $S_3$ (a similar construction is given in Ref. \onlinecite{TuChang21}). We represent $N=\inner{r|r^3=1}$ and $Q=\inner{s|s^2=1}$. The group extension is a split extension whose factor system is given by $\sigma^{s}[r]=\bar r$ and trivial cocycle $\omega=1$.

The operators on each edge can be represented by $\ZZ_2$ Pauli operators, while those on each plaquette are represented by $\ZZ_3$ clock and shift matrices $\cZ$ and $\cX$, which satisfy $\cZ \cX = e^{2\pi i /3}\cX \cZ$. The automorphism operator acts as $\Sigma^s_p = \cC_p$ where $\cC$ is the $\ZZ_3$ charge conjugation operator
\begin{align}
\cC = \begin{pmatrix}
1 &0 &0\\
0&0&1\\
0&1&0
\end{pmatrix}
\end{align}

Starting with $\bs A^g_v$, we can substitute
\begin{align}
    L_p^r = R_p^{\bar r}&\equiv \cX_p \\
     R_p^r = L_p^{\bar r}&\equiv \bar \cX_p\\
    R_e^s = L_e^s &\equiv X_e\\
    L_p^{\Sigma^e[r]}  = R_p^{\Sigma^e[\bar r]}&\equiv \cX_p^{Z_e} \\
    L_p^{\Sigma^e[\bar r]} = R_p^{\Sigma^e[r]}& \equiv \bar \cX_p^{Z_e}
\end{align}

With this, we find
\begin{align}
    \bs A^{\fs{1}{s}}_v =&  \prod_{e \supset v} X_e \prod_{p|v=i} \cC_p \\
\bs A^{\fs{r}{1}}_v=&\prod_{p|v=j,l} \bar \cX_p^{Z_{iv}} \prod_{p|v=k} \cX_p^{Z_{iv}} \prod_{p|v=i} \cX_p
\end{align}

The projector $\bs B^Q_{\nablapic}$ is the same as the previous case
\begin{align}
    \bs B^Q_{\nablapic} = \frac{1 + \prod_{e \subset \nablapic}Z_e}{2}
\end{align}

Lastly, the constraint of the cube operator $\bs B_{c,x}$ enforces
\begin{align}
   \aut{q_{ac}}{n_y} \aut{q_{ad}}{\bar n_{z}} \bar n_{\bar y} n_{\bar z}=1
\end{align}
Define the operator
\begin{align}
    \bs B_{c,x}^{r} = \cZ_y^{Z_{ac}} \bar \cZ_{z}^{Z_{ad}} \bar \cZ_{\bar y} \cZ_{\bar z}
\end{align}
which satisfies the constraint precisely when its eigenvalue is one. Therefore, the projector can be written as
\begin{align}
    \bs B_{c,x}^N = \frac{1+\bs B_{c,x}^{r} + \left (\bs B_{c,x}^{r}\right)^2}{3}
\end{align}

\subsection{$(D_n,\ZZ_n)$}
The example of $(D_{n},\ZZ_n)$ is naturally obtained by generalizing the plaquette degrees of freedom in the $(S_3,\ZZ_3)$ theory above to $\ZZ_n$ qudits. Namely, the clock and shift matrices now satisfy $\cZ \cX = e^{2\pi i /n}\cX \cZ$, and the projector $\bs B_{c,x}^N$ is now given by
\begin{align}
    \bs B_{c,x}^N = \frac{1}{n}\sum_{m=1}^{n} (\bs B_{c,x}^{r})^m
\end{align}

The model can be thought of as starting with the $\ZZ_n$ X-cube model with degrees of freedom on the plaquettes, and gauging the charge conjugation symmetry, which results in a $\ZZ_2$ gauge field living on the edges.

\subsection{$(Q_8,\ZZ_4)$}
As a final example, we present the simplest non-abelian hybrid fracton model where the corresponding group extension is neither central nor split. The smallest group where this happens is $G=Q_8 = \inner{r,s|r^4=r^2s^2=(rs)^2=1}$ the quaternion group, $N=\ZZ_4 = \inner{r|r^4=1}$, and $Q=\ZZ_2 = \inner{s|s^2=1}$. The factor system is given by $\sigma^s[r]=\bar r$ and $\omega(s,s)=r^2$.

Each edge can be expressed using $\ZZ_2$ Pauli operators, while each plaquette can be expressed with $\ZZ_4$ clock and shift operators, which satisfy $\cZ\ \cX = i \cX \cZ$. The automorphism operators acts as $\Sigma^s_p = \cC_p$, where $\cC$ is the $\ZZ_4$ charge conjugation operator
\begin{align}
  \cC = \begin{pmatrix}
1 &0 &0 &0\\
0&0&0&1\\
0&0&1&0\\
0&1&0&0
\end{pmatrix}  
\end{align}

The substitution to Pauli operators for the $\bs A^g_v$ terms are
\begin{align}
    L_p^r = R_p^{\bar r}&\equiv \cX_p \\
     R_p^r = L_p^{\bar r}&\equiv \bar \cX_p\\
    R_e^s = L_e^s &\equiv X_e\\
    L_p^{\Sigma^e[r]}  = R_p^{\Sigma^e[\bar r]}&\equiv \cX_p^{Z_e} \\
    L_p^{\Sigma^e[\bar r]} = R_p^{\Sigma^e[r]}& \equiv \bar \cX_p^{Z_e}\\
    L_p^{\Omega(e,s)}=R_p^{\Omega(e,s)} &\equiv \left (\cX_p^2\right)^\frac{1-Z_e}{2} = \cX_p^{1-Z_e}
\end{align}
Therefore, we have
\begin{align}
\bs  A^{\fs{1}{s}}_v  =&\prod_{p|v=j,k,l}  \cX_p^{1-Z_{iv}}  \times \prod_{e \supset v } X_e \times \prod_{p|v=i} \cX_p^{Z_{vj}-Z_{vk}+Z_{vl}-1} \cC_p.\\
\bs A^{\fs{r}{1}}_v=&\prod_{p|v=j,l} \bar \cX_p^{Z_{iv}} \prod_{p|v=k} \cX_p^{Z_{iv}} \prod_{p|v=i} \cX_p
\end{align}
The projector $\bs B^Q_{\nablapic}$ is again
\begin{align}
    \bs B^Q_{\nablapic} = \frac{1 + \prod_{e \subset \nablapic}Z_e}{2}
\end{align}

Lastly, to express $\bs B^N_{c,x}$ we define
\begin{align}
    \bs B^r_{c,x} =& CZ_{ad,ds} CZ_{ad,dw} CZ_{ad,du} CZ_{ac,ct} CZ_{ac,cw} CZ_{ac,cu} \nonumber \\
     &\cZ_y^{Z_{ac}} \bar \cZ_{z}^{Z_{ad}} \bar \cZ_{\bar y} \cZ_{\bar z}
\end{align}
which enforces the constraint when $\bs B^r_{c,x}=1$. Therefore, the projector is given by
\begin{align}
\bs B^N_{c,x} = \frac{1+\bs B^r_{c,x} + (\bs B^r_{c,x})^2 + (\bs B^r_{c,x})^3}{4}
\end{align}

The above construction can be straightforwardly generalized to any dicyclic group $(Q_{4n},\ZZ_{2n})$ by replacing each plaquette with $\ZZ_{2n}$ qudits.

\bibliography{references}
\end{document}